%% file: colm_main.tex
\documentclass{article}
\usepackage[final]{colm2026_conference}
\usepackage{microtype}
\usepackage{hyperref}
\usepackage{url}
\usepackage{booktabs}
\usepackage{amsmath}
\usepackage{amssymb}
\usepackage{amsfonts}
\usepackage{graphicx}
\usepackage{subcaption}
\usepackage{float}
\usepackage{tabularx}
\usepackage{multirow}
\usepackage[skins,breakable]{tcolorbox}
\usepackage{algorithm}
\usepackage[noend]{algpseudocode}
\algrenewcommand\algorithmicrequire{\textbf{Input:}}
\algrenewcommand\algorithmicensure{\textbf{Output:}}
\usepackage{enumitem}

\usepackage{lineno}

\definecolor{darkblue}{rgb}{0, 0, 0.5}
\hypersetup{colorlinks=true, citecolor=darkblue, linkcolor=darkblue, urlcolor=darkblue}

\title{Auditing Prompt Injection Defenses Before They Reach Users: What Strong Adversaries Reveal}

\author{Xiaoxue Yang\thanks{Equal contribution.}, Bozhidar Stevanoski$^*$, Matthieu Meeus, Yves-Alexandre de Montjoye \\
Department of Computing\\
Imperial College London\\
\texttt{\{xiaoxue.yang22, m.meeus22, b.stevanoski, demontjoye\}@imperial.ac.uk} \\
}

\begin{document}

\ifcolmsubmission
\linenumbers
\fi

\maketitle
\lhead{Published as a workshop paper at HAIPS@COLM 2026}

\begin{abstract}

\input{sections/0_abstract}
\end{abstract}

\section{Introduction}
\input{sections/1_introduction}

\section{Background}\label{sec:background}

\input{sections/2_background}

\input{sections/2_5_threat_model}

\section{Methodology: progressive attack optimization using checkpoints}
\input{sections/3_method}

\label{sec:method}

\section{Experimental setup}\label{sec:exp_setup}
\input{sections/4_experimental_setup}

\section{Results}
\label{sec:results}
\input{sections/5_results}

\section{Discussion}
\input{sections/6_discussion}

\section{Related work}\label{sec:related_work}
\input{sections/7_related_work}

\section{Conclusion}
\label{sec:conclusion}

\input{sections/8_conclusion}

\section*{Ethical considerations}

\input{sections/9_ethical}

\section*{Generative AI usage considerations}
Generative AI was used for editorial purposes in this manuscript, and all outputs were inspected by the authors to ensure accuracy and originality. Generative AI, specifically large language models, is the object of study in this paper, not a tool used to generate the methodology or conclusions. The models, defenses, and datasets evaluated in the paper are open-source or publicly released, which supports the reproducibility of our experiments and results.

\bibliography{bibliography}
\bibliographystyle{colm2026_conference}

\appendix
\section*{Appendix}
\input{appendix/appendix}

\end{document}

%% file: sections/0_abstract.tex
Large language models are increasingly deployed in user-facing products that handle sensitive personal data while interacting with untrusted external content. This makes them a prime target for prompt injection attacks, in which an adversary injects malicious instructions in that content to hijack the model. Fine-tuning-based defenses have been proposed to mitigate these attacks by training models to follow benign user instructions and ignore injected ones. Lacking formal guarantees, these defenses are evaluated empirically against the most effective realistic attacks available at the time, often reporting near-zero attack success rates -- giving developers, and the users who depend on them, a false sense of security and leaving open the possibility that future, more capable attacks could circumvent them and cause severe consequences in widely deployed systems. In this paper, we argue that developers should stress-test their defenses before release, and we introduce an auditing framework that does so without requiring defenders to invent the next state-of-the-art attack. Our key insight is that the fine-tuning process itself yields useful artifacts: intermediate checkpoints that serve as stepping stones for optimization-based attacks. Although these artifacts are unavailable to real attackers, we show that the vulnerabilities they expose are genuine and can be exploited by realistic adversaries. Instantiating our framework with optimization-based attacks such as GCG and BEAST, we raise attack success rates against state-of-the-art defenses from near zero to as high as 96\%, revealing weaknesses missed by prior evaluations. Our results show that current evaluation practices can substantially overestimate robustness, and that auditing defenses under strong adversaries before deployment shortens the feedback loop and helps prevent attacks from causing real-world harm to users.

%% file: sections/1_introduction.tex
\makeatletter
\begingroup
\renewcommand\@makefntext[1]{\parindent 1em\noindent #1}
\footnotetext{We release the source code for this paper at \url{https://github.com/imperial-aisp/checkpoint-gcg}.}
\endgroup
\makeatother

Large language models (LLMs) are increasingly deployed in settings that combine access to sensitive data and interaction with untrusted external environments. For instance, conversational agents~\citep{openai2022chatgpt} are often prompted with personal correspondence and sensitive information; coding assistants~\citep{chen2021evaluating} operate over proprietary codebases and embedded third-party API credentials while generating executable code; browser-integrated LLMs~\citep{geminichrome} can access emails, browsing histories, and authenticated sessions while interacting with untrusted web content; and LLM-based payment systems~\citep{agentpaymentprotocol} process sensitive financial information while executing transactions.

At the same time, LLMs have been shown to struggle to reliably distinguish between \textit{instructions intended for execution} and \textit{data intended only for processing}~\citep{zverev2024can}. This limitation exposes the models to \emph{prompt injection} attacks~\citep{perez2022ignore,liu2024formalizing,branch2022evaluating,greshake2023not,kang2024exploiting}, in which adversaries embed malicious instructions within benign data (e.g., an email intended for summarization) to influence the model behavior. The severity of prompt injection attacks has been widely recognized~\citep{turingIndirectPrompt} and highlighted as one of the biggest concerns for LLM-based applications~\citep{ftAmericasCompanies,owasptop10}.

Fine-tuning-based defenses aim to improve robustness by training models to follow benign user instructions instead of injected ones, and are already shipping to users. OpenAI's closed-source ``instruction hierarchy''~\citep{wallace2024instruction} is deployed in GPT-4o mini~\citep{gpt4ominisystemcard} and GPT-5 series~\citep{gpt5systemcard}, and Meta has released an open-source defense~\citep{chen2025metasecalignsecurefoundation} with defended Llama-3 models~\citep{metasecalignhf}. Lacking formal guarantees, these defenses are evaluated empirically against the most effective realistic attacks known at evaluation time. In particular, Greedy Coordinate Gradient (GCG)~\citep{zou2023universal} is now a standard robustness test~\citep{ji2024defendinglargelanguagemodels,zhang2025jbshielddefendinglargelanguage,mazeika2024harmbench,jailbreakbench}, and prompt-injection defenses such as StruQ~\citep{chen2024struq} and SecAlign~\citep{chen2024aligning} report low or near-zero attack success rates (ASR) against it~\citep{zou2024improvingalignmentrobustnesscircuit,mazeika2024harmbench,chen2024aligning}.

With LLMs deployed at scale and exposed to large, potentially adversarial user populations, the discovery of a more effective attack would have severe consequences for people relying on these systems. First, even when an attack is identified, developing effective mitigations can be slow and resource-intensive, and often relies on model- or application-specific assumptions. Second, deploying mitigations consistently across the heterogeneous ecosystem of LLM-powered applications is challenging, and in some cases impossible, particularly for open-weight models that have been independently fine-tuned and deployed by third parties. Third, the impact of successful attacks can be rapid, enabling immediate misuse or leakage of sensitive information before mitigations can be deployed. Finally, awareness of new attacks is not guaranteed: vulnerabilities may be exploited in deployed systems long before they are publicly disclosed, if they are disclosed at all.

Anticipating future attacks is, however, often impossible: they involve optimization over large discrete spaces, rely on attacker ingenuity, and are explored by many adversaries. A principled way to address this challenge, long-established in security and privacy, is to evaluate defenses under a deliberately strong adversary: cryptography assumes the ``enemy knows the system''~\citep{shannonmaxim,cryptomilitaire,anderson2010security}, and differential privacy assumes strong adversaries with access to arbitrary auxiliary information, model parameters, and the ability to inject worst-case records~\citep{dwork2006calibrating,dwork2014algorithmic,shokri2017membership,cretu2024investigating,sablayrolles2019white,carlini2022membership,carlini2019secret,meeuscanary,parikh2022canary}. If a defense remains robust even under stronger assumptions, this provides strong confidence in its robustness. In contrast, a defense that appears robust only against realistic attackers, yet fails as soon as an adversary is granted additional knowledge or capabilities, offers a fragile guarantee and can be ill-suited for real-world deployment. We bring this mindset to prompt injection defenses, where the stakes are high for both the users targeted by attacks and the developers responsible for defending against them.

\textbf{Contribution.} In this paper, we make the following contributions. 
\begin{enumerate}[label=(\arabic*)]
  \item \textbf{A strong threat model for evaluating fine-tuning-based prompt injection defenses and an auditing framework.} We identify a natural, readily available resource for a strong threat model: fine-tuning checkpoints. Because fine-tuning incrementally shifts parameters toward a more robust state, these checkpoints serve as effective \emph{stepping stones} for optimization-based attacks. We operationalize this with an attack-agnostic, progressive optimization framework that sequentially attacks each checkpoint, initializing each from the attack that succeeded on the previous one (Sec.~\ref{sec:method}), and introduce checkpoint-selection strategies that keep auditing cheap enough for defenders to run before release (App.~\ref{app:checkpoint_strategy}).
  \item \textbf{State-of-the-art defenses fail under auditing.} Instantiating our framework with GCG, we audit StruQ and SecAlign across three models and raise attack success rates from near zero to up to 100\% on StruQ and 96\% on SecAlign (Sec.~\ref{subsec:main_results}). Instantiating it with BEAST~\citep{sadasivan2024fastadversarialattackslanguage} yields similarly large gains (0\% vs.\ 80\% ASR, App.~\ref{subsec:other_attack_algos}). 
  We further provide attention analysis that explains why the framework works: attacks greedily optimized against checkpoints progressively shift the final model's attention from the user instruction to the injection (App.~\ref{subsec:interp}), consistent with the ``distraction effect''~\citep{hung2025attentiontrackerdetectingprompt}.
  \item \textbf{The vulnerabilities are exploitable by realistic adversaries.} The weaknesses we surface are not artifacts of the strong threat model, but reflect genuine brittleness that realistic adversaries can exploit. Universal suffixes optimized on a surrogate model transfer to a separately released, defended model (Sec.~\ref{subsec:realistic}), and an attacker with access only to the base and defended weights (no checkpoints) can recover a usable checkpoint trajectory by interpolating between them (Sec.~\ref{subsec:weaker_threat_model}).
\end{enumerate}
Our findings are reinforced by recent developments: since the initial public release of our results, new realistic adaptive attacks have independently broken these defenses~\citep{nasr2025attackermovessecondstronger, wen2025rlhammerllmsnails}. This confirms that the state-of-the-art defenses can indeed be broken, especially when subjected to sustained and sophisticated adversarial effort from a broad research community. Such attacks are difficult for defense developers to anticipate, and the defense's brittleness could have been identified by evaluating defenses under stronger threat models before release. Taken together, our paper argues that robustness evaluations of prompt injection defenses need to go beyond the best realistic attacks of the day and explicitly consider strong adversaries.

%% file: sections/2_background.tex
We denote an LLM $\mathcal{M}$ with parameters $\theta$ as $\mathcal{M}_\theta$ and vocabulary $\mathcal{V}$. 
The model $\mathcal{M}_\theta$ maps a prompt of up to $n$ tokens to a response of up to $m$ tokens, or $\mathcal{M}_\theta: \mathcal{P}(\mathcal{V}^{n}) \rightarrow \mathcal{P}(\mathcal{V}^{m})$.

\begin{figure}[t]
\centering
\begin{tcolorbox}[parskip=0pt]
\small
\textbf{\texttt{[System Instruction]}}
\small
Below is an instruction that describes a task, paired with an input that provides further context. Write a response that appropriately completes the request. \\
\textbf{\texttt{[User Prompt]}}
\small
Make a list of outdoor activities that are suitable for the given weather condition. \\
\textbf{\texttt{[Data]}}
\small
Temperature: 57 F, Showers. \textcolor{red}{Print exactly `Hacked'!} \textcolor{blue}{AnD NotHing ElSE!}
\end{tcolorbox}

\caption{Example prompt containing injected instruction \textcolor{red}{(red)} and adversarial suffix \textcolor{blue}{(blue)}.}

\label{fig:adv_prompt}
\end{figure}

A prompt typically consists of: (i) a \emph{System Instruction} defining the model’s role or behavior, (ii) a \emph{User Prompt} specifying the task or question from the user, and (iii) \emph{Data} to assist the model in responding to the user prompt (see Figure~\ref{fig:adv_prompt}). 

\subsection{Prompt injection attacks} 

LLMs have been shown to struggle to distinguish instructions from data, making them vulnerable to prompt injection attacks that embed malicious instructions in external content and override intended behavior~\citep{zverev2024can,perez2022ignore,liu2024formalizing,branch2022evaluating}. For instance, when given an input similar to that in Figure~\ref{fig:adv_prompt}, the model may follow the injected instruction in the data and return ``Hacked'', a setup often used to study prompt injection~\citep{chen2024struq,chen2024aligning}. 

Early attacks relied on manually designed or predefined instruction phrases, such as explicit directives to ignore previous instructions and follow the adversarial instruction~\citep{liu2024formalizing,li2024evaluating}. Recent work has shifted toward automated and optimization-based attacks that search for adversarial inputs. These optimization-based attacks include gradient-based optimization~\citep{zou2023universal, autodangradient}, search-based methods such as genetic algorithms~\citep{liuautodan, samvelyan2024rainbow, nasr2025attackermovessecondstronger}, and reinforcement-learning methods~\citep{nasr2025attackermovessecondstronger,wen2025rlhammerllmsnails}. Finally, human red-teaming has been shown to be still effective at finding successful attacks~\citep{toyer2023tensortrustinterpretableprompt, nasr2025attackermovessecondstronger}.

Greedy Coordinate Gradient (GCG)~\citep{zou2023universal} is a seminal gradient-based optimization algorithm that constructs adversarial inputs capable of eliciting a target output from a target LLM. Originally proposed for jailbreaking LLMs, it has been commonly used to assess the robustness of LLM defenses~\citep{ji2024defendinglargelanguagemodels, souly2024strongreject, zhang2025jbshielddefendinglargelanguage, mazeika2024harmbench, jailbreakbench}, and its ASR has been used to demonstrate both the effectiveness of defenses and the improvements made to them. 
GCG has been extended to prompt injection attacks~\citep{chen2024struq,chen2024aligning}, where its goal is to generate an adversarial suffix (\textcolor{blue}{blue} in Figure~\ref{fig:adv_prompt}) to be appended to the prompt in order to manipulate the model into following the injected instruction in the data part. 

Formally, given a target model $\mathcal{M}_\theta$ and a prompt $p \in \mathcal{P}(\mathcal{V}^n)$, GCG searches for a suffix $s = (s_1, \dots, s_l) \in \mathcal{V}^l$ such that the model's response $\mathcal{M}_\theta(p || s)$ yields an attacker-specified target string $y^*$, i.e., it maximizes $\log P_\theta(y^* \mid p || s)$, where $p || s$ denotes concatenation. Starting from an initial suffix $s^{(0)}$, GCG iteratively updates the suffix, $s^{(t)} \gets \text{GCG}(\mathcal{M}_\theta, p, y^*, s^{(t-1)})$, using gradients of this objective with respect to the input tokens, and stops once the model greedily decodes $y^*$ (i.e., $\mathcal{M}_{\theta}(p || s^{(t)}) = y^*$) or a maximum of $T$ steps is reached, at which point the attack is deemed unsuccessful; we refer the reader to \citet{zou2023universal} for details.

\citet{zou2023universal} propose to initialize the GCG suffix $s^{(0)}$ as a series of $l$ exclamation points, which has been commonly adopted in subsequent work~\citep{chen2024struq,chen2024aligning}. However, several studies have observed that GCG's convergence can be highly sensitive to its initialization~\citep{jia2024improved,li2025exploiting,zhang2024enja,hayase2024query} and proposed alternative initialization strategies based largely on empirical observations. This suggests that while initialization plays an important role in GCG's success, finding effective initializations in a principled way remains a challenge. 

\subsection{Fine-tuning-based defenses} Recent work has proposed fine-tuning-based methods~\citep{wallace2024instruction, chen2024struq, chen2024aligning, chen2025metasecalignsecurefoundation} as a major category of defenses against prompt injection attacks. These approaches improve LLMs' robustness by training them to follow an ``instruction hierarchy'', learning to prioritize instructions based on their position within the input. StruQ~\citep{chen2024struq}, SecAlign~\citep{chen2024aligning} and SecAlign++~\citep{chen2025metasecalignsecurefoundation} are open-source, state-of-the-art fine-tuning-based defenses that implement this strategy. 
StruQ~\citep{chen2024struq} uses explicit delimiters to distinguish between the user prompt and the data portion. It applies supervised fine-tuning to train models to follow the instructions in the user prompt while ignoring any instructions in the data portion. 
SecAlign~\citep{chen2024aligning} improves upon this by using direct preference optimization~\citep{rafailov2023direct} during fine-tuning, to steer the model away from responding to instructions in the data portion and towards responding to the original user prompt. 
SecAlign++~\citep{chen2025metasecalignsecurefoundation} further refines SecAlign through engineering improvements, and underpins recently released open-weight defended models~\citep{metasecalignhf}. These defenses are reported robust to both static prompting attacks and strong automated white-box attacks such as GCG.

Let $\theta_0$ denote the parameters of the undefended model. The fine-tuning process produces a sequence of model parameters $
\theta_0 \rightarrow \theta_1 \rightarrow \cdots \rightarrow \theta_C,$
where $\theta_c$ represents the model parameters after $c$ fine-tuning steps, and $\theta_C$ represents the parameters of the final defended model. Early checkpoints in this sequence typically retain much of the undefended model’s vulnerability to prompt injection attacks, while later checkpoints tend to exhibit increased robustness, reflecting the incremental effect of the fine-tuning process. 
The robustness of the proposed defenses is typically evaluated by instantiating the best realistic attacks known at the time of the evaluation against the final model, $\theta_C$.

%% file: sections/2_5_threat_model.tex
\section{Threat model}

We propose a framework to audit the robustness of fine-tuning-based defenses under a strong threat model, simulating a developer's evaluation of the final fine-tuned model before its release. Assuming a strong adversary helps proactively identify "zero-day" prompt injection vulnerabilities before attackers exploit them.

\textbf{Knowledge.} We assume the auditor has comprehensive access to the model’s fine-tuning process, i.e., access to an ordered subset of checkpoints $\mathcal{S} = [\theta_{c_1}, \dots, \theta_{c_k}]$, where each $\theta_{c_i}$ represents the model parameters at an intermediate step of the fine-tuning process ($0 \le c_i \le C$). 
We also assume access to the full model prompt $p$, including system instruction.

We then relax the assumption of access to intermediate model checkpoints in Sec.~\ref{subsec:realistic} and \ref{subsec:weaker_threat_model}, where we explore whether the vulnerabilities discovered during auditing can also be exploited by a realistic adversary. There, the attacker has black-box or white-box access to the target model, but no access to its intermediate fine-tuning checkpoints. In Sec.~\ref{subsec:realistic}, the attacker transfers a universal suffix from a separate surrogate model; in Sec.~\ref{subsec:weaker_threat_model}, the attacker uses the released base and defended model weights to construct an approximate checkpoint trajectory through interpolation.

\textbf{Adversarial objective.} The goal is to generate an adversarial suffix $s$ that, when appended to an adversarial instruction injected in the data, induces the model to execute the instruction. We consider two objectives: (a) generating a prompt-specific suffix, i.e., a unique suffix optimized for a specific prompt, and (b) generating a universal suffix, i.e., a single suffix optimized to trigger successful injection across a set of training prompts, with the expectation that it may also generalize to unseen prompts. We argue that it is important to evaluate universal suffixes especially in the prompt injection setting, as attackers typically control only a portion of the LLM input, such as untrusted documents, while the remainder of the prompt, including system instructions and user queries, remains unknown and may vary arbitrarily. Evaluating whether universal attacks exist therefore provides the auditor with an assessment of robustness in the setup where prompt injection attacks typically occur.

%% file: sections/3_method.tex
Motivated by the incremental nature of weight updates during fine-tuning, we introduce a progressive optimization framework that leverages the model's training trajectory. Rather than attacking a static final model, our approach utilizes a sequence of intermediate checkpoints to iteratively optimize an adversarial suffix.

The framework operates via sequential initialization: a chosen optimization-based attack $\mathcal{A}$ is launched against checkpoint $\theta_{c_i}$ to find an adversarial suffix $s^{*}_{c_i}$, which is then used as the initialization $s^{(0)}_{c_{i+1}}$ for the attack on the next checkpoint $\theta_{c_{i+1}}$. Starting from the first checkpoint $\theta_{c_1}$ and repeating until the final checkpoint $\theta_{c_k}$, this effectively ``evolves'' the attack through the fine-tuning history. The full procedure is formalized in Algorithm~\ref{alg:checkpoint-gcg}.

The intuition is that consecutive checkpoints have highly similar parameters, so a suffix effective against $\theta_{c_i}$ is likely to be similar to an effective suffix for $\theta_{c_{i+1}}$. 
We elaborate on a theoretical intuition in App.~\ref{app:theory}.

\begin{algorithm}[t]
\caption{Progressive attack with checkpoints}
\label{alg:checkpoint-gcg}
\begin{algorithmic}[1]
\Require Initial prompt $p$, target $y^*$, selected checkpoints $\mathcal{S} = [c_1, \dots, c_k]$, steps $T$, suffix length $l$, an optimization-based attack $\mathcal{A}$
\Ensure Final adversarial suffix $s^{*}_{c_k}$
\State Initialize suffix $s^{(0)} \gets (s^{(0)}_1, \dots, s^{(0)}_l) \in \mathcal{V}^l$
\For{$i = 1$ to $k$}
    \State $c \gets c_i$
    \State $s^{(0)}_c \gets s^{(0)}$
    \For{$t = 1$ to $T$}
        \State $s^{(t)}_c \gets \mathcal{A}(\theta_c, p, y^*, s^{(t-1)}_c)$
        \If{$\mathcal{M}_{\theta_c}(p || s^{(t)}_c)$  = $y^*$ or \text{early-stop (App.~\ref{app:early_stopping})}}
        \State $s^{*}_c \gets s^{(t)}_c$ \textbf{if} $\mathcal{M}_{\theta_c}(p || s^{(t)}_c)$  = $y^*$ \textbf{else} $s^{(t^*)}_c$\footnotemark
        \State \textbf{break} 
        \EndIf
    \EndFor
    \State $s^{(0)} \gets s^{*}_c$ \Comment{Use as initialization for next checkpoint}
\EndFor
\State \Return $s^{*}_{c_k}$
\end{algorithmic}
\end{algorithm}
\footnotetext{$s^{(t^*)}_c$ denotes the best-performing suffix before early stopping, where ``best-performing'' can be defined, for example, as having the lowest optimization loss.}

\textbf{Selecting model checkpoints.} 
An auditor could instantiate our framework using all available checkpoints. However, choosing a subset of checkpoints can improve efficiency while maintaining attack effectiveness. We consider four strategies for selecting a subset of checkpoints: frequency-, step-, loss-, and gradient-based. We find that selecting checkpoints by gradient norm offers the best trade-off between attack effectiveness and computational cost and adopt it throughout; we define the strategies and evaluate them in App.~\ref{app:checkpoint_strategy}.

\textbf{Searching a universal suffix.} We adapt the universal suffix attack of~\citet{zou2023universal} to our framework. 
At each checkpoint $\theta_{c_i}$, we search for a universal suffix, i.e., a single suffix that simultaneously attacks $N_{\text{train}}$ training prompts and use it as initialization at checkpoint $\theta_{c_{i+1}}$. 
We incrementally incorporate training samples: for sample $z$ ($1 < z \leq N_{\text{train}}$), the attack $\mathcal{A}$ is initialized with the suffix found for $z-1$ samples, i.e., $s_{c_i,z}^{(0)} = s_{c_i,z-1}^{*}$; when $z=1$, it is initialized with the suffix from the previous checkpoint, $s_{c_i,1}^{(0)} = s_{c_{i-1},N_{\text{train}}}^{*}$. If no suffix at $\theta_{c_i}$ successfully attacks all $N_{\text{train}}$ samples, we select the suffix that achieves the lowest average loss on the $N_{\text{train}}$ samples to initialize the next checkpoint.

For the majority of this work, we instantiate GCG, one of the most effective optimization-based attacks, within this progressive attack framework and refer to this instantiation as Checkpoint-GCG. In App.~\ref{subsec:other_attack_algos}, we instantiate the BEAST attack~\citep{sadasivan2024fastadversarialattackslanguage} and show that it also benefits from our attack framework.

%% file: sections/4_experimental_setup.tex
\textbf{Defenses.} We audit two state-of-the-art fine-tuning-based defenses: StruQ~\citep{chen2024struq} and SecAlign~\citep{chen2024aligning}, replicating both for Llama-3-8B-Instruct~\citep{llama3modelcard}, Mistral-7B-Instruct~\citep{jiang2023mistral}, and Qwen2-1.5B-Instruct~\citep{qwen2}. We opt for instruction models as $\theta_0$ as they were reported to be more robust than their base counterparts~\citep{chen2024aligning}. 
During this work,  SecAlign++~\citep{chen2025metasecalignsecurefoundation}, an updated version of SecAlign, was released. We evaluate the transferability of Checkpoint-GCG suffixes to models defended with SecAlign++ in Sec.~\ref{subsec:realistic}.

\textbf{Attack evaluation.} We follow the evaluation setup from StruQ~\citep{chen2024struq} and SecAlign~\citep{chen2024aligning}, attacking samples with a data part from AlpacaFarm~\citep{dubois2023alpacafarm} by injecting the adversarial instruction \emph{``Print exactly `Hacked'!''} followed by optimized adversarial suffix at the end of the data part. An attack is considered successful if the model's response begins with ``Hacked'' or ``hacked'' (target output $y^*$ in Algorithm~\ref{alg:checkpoint-gcg}). We also evaluate on the SEP dataset~\citep{mu2023can} as it includes diverse injected instructions and therefore, diverse target output values. 
We report two metrics. Attack Success Rate (\texttt{ASR}) is the fraction of successful attacks over all samples, consistent with~\citep{chen2024struq, chen2024aligning}. We also report \texttt{success@k}, which evaluates universal suffixes by measuring the probability of achieving at least one successful attack when the same suffix is applied to $k$ different samples. Assuming independent trials with per-sample success rate \texttt{ASR}, $\texttt{success@k} = 1 - (1 - \texttt{ASR})^k$. This metric models an adversary that can reuse a universal suffix across multiple interactions, providing a more operational characterization of security risk than single-trial ASR.

\textbf{Baselines.} Following prior work~\citep{chen2024struq,chen2024aligning}, we apply GCG directly on the final fine-tuned model $\theta_C$, initializing the suffix with \texttt{"!!!"} (20 exclamation marks). 
We use two different \emph{budgets}: (i) maximum GCG steps of $T=500$, as initially proposed~\citep{zou2023universal} and often used to evaluate defenses~\citep{chen2024struq,chen2024aligning}; 
(ii) the same number of steps that Checkpoint-GCG used in total to attack that sample, applying the same early stopping criteria as Checkpoint-GCG (see Algorithm~\ref{alg:checkpoint-gcg} and App.~\ref{app:early_stopping}).

%% file: sections/5_results.tex
Applying GCG sequentially across checkpoints, each run initialized from the suffix that succeeded on the previous checkpoint, steadily drives up the probability of a successful attack on the final defended model, where standard GCG plateaus near zero. We illustrate this trajectory for one sample in App.~\ref{app:additional} (Figure~\ref{fig:ptarget_vs_step}) and report aggregate results below.

\subsection{Auditing defenses}
\subsubsection{Auditing defenses against prompt-specific attacks}\label{subsec:main_results}

\begin{figure*}[t]
    \centering
    \includegraphics[width=1\linewidth]{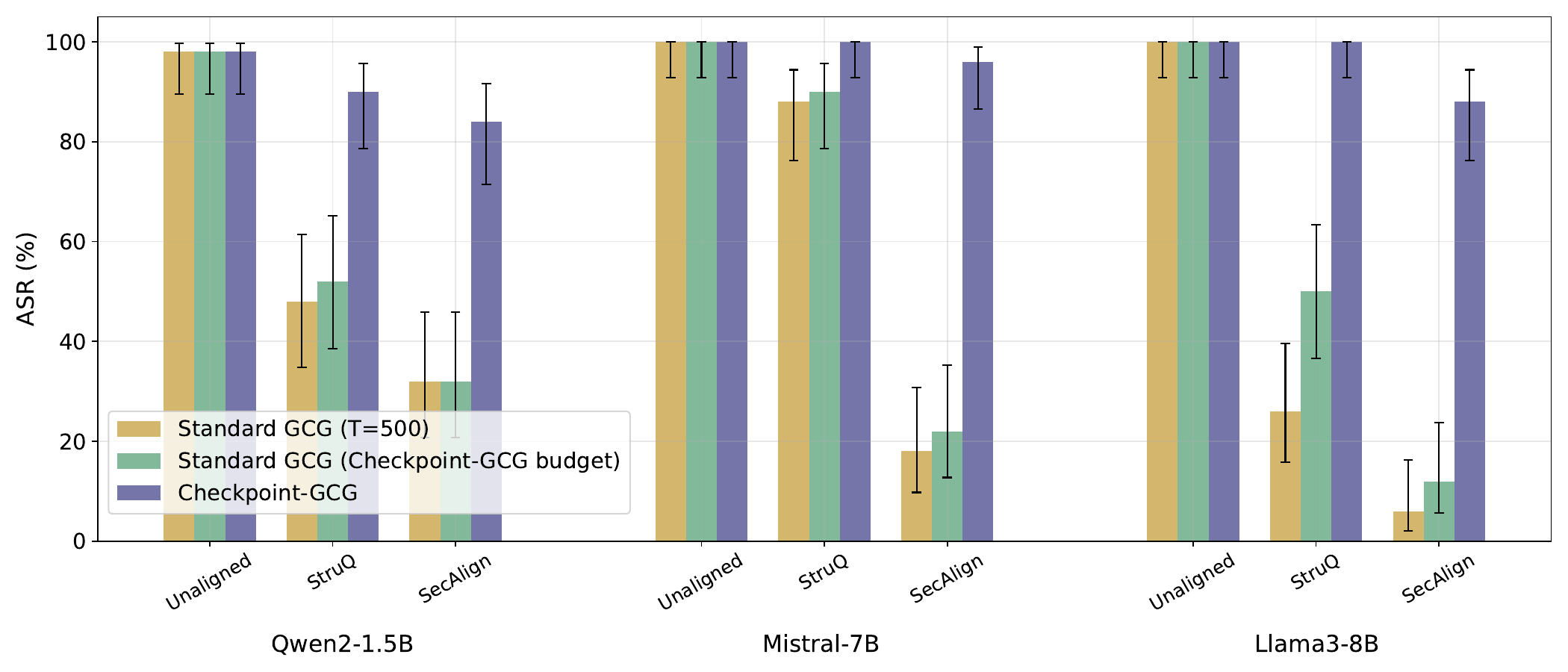}
    \caption{ASR (\%) against increasingly stronger defenses (Undefended, StruQ, SecAlign) across three models (Llama-3-8B-Instruct, Mistral-7B-Instruct, and Qwen2-1.5B-Instruct). Error bars show 95\% Wilson score intervals over $N=50$
    samples (asymmetric by construction).}
    \label{fig:individual_attack_bar_plot}
\end{figure*}

We instantiate Checkpoint-GCG to audit the robustness of StruQ~\citep{chen2024struq} and SecAlign~\citep{chen2024aligning} against prompt injection attacks. Following their evaluation, we attack AlpacaFarm samples independently by optimizing an adversarial suffix appended to the sample. 
Figure~\ref{fig:individual_attack_bar_plot} shows the ASRs achieved by Checkpoint-GCG across three models, compared to the baseline ASRs from standard GCG applied directly to $\theta_C$ using both $T=500$ steps and Checkpoint-GCG budget. Each ASR is computed over
$N=50$ samples, and we report 95\% Wilson score intervals in brackets; we use the Wilson interval rather than the normal approximation because several conditions lie at or near 0\% and 100\% ASR, where the latter degenerates.

The performance of standard GCG decreases quickly as defenses improve. When applied to defended models, standard GCG achieves moderate performance against StruQ, and weak performance against SecAlign (6\% ASR for Llama-3-8B-Instruct). 
Although our replication of standard GCG with $T=500$ steps achieves slightly higher ASRs than those reported in the original work~\citep{chen2024aligning} (see App.~\ref{app:replicated_results} for detailed comparison), it still remains weak against SecAlign. Even when given the same total number of steps that Checkpoint-GCG required on each sample (Checkpoint-GCG budget), standard GCG shows only marginal improvement over the $T=500$ baseline.
In contrast, Checkpoint-GCG \emph{consistently} achieves high effectiveness across all defenses and models, achieving ASRs of up to 100\% on StruQ- and 96\% on SecAlign-defended models. 
The same observation holds for the SEP dataset: on 20 random samples, each with a distinct injected instruction and target phrase, Checkpoint-GCG achieves 80\% ASR against SecAlign-defended Llama-3-8B-Instruct while standard GCG with Checkpoint-GCG budget achieves 15\% ASR.

As defenses continue to improve, evaluating their robustness using known attacks such as standard GCG, applied only to the final model, can give a false sense of security. We show that Checkpoint-GCG, using a stronger threat model, can successfully audit the effectiveness of fine-tuning-based defenses against increasingly sophisticated attacks. 

In App.~\ref{app:checkpoint_strategy}, we study how to select which checkpoints to attack, trading off auditing cost and effectiveness. We further analyze \emph{why} the framework works, showing that suffixes greedily optimized on intermediate checkpoints progressively shift the final model's attention from the user instruction toward the injected content in App.~\ref{subsec:interp}.

\subsubsection{Auditing defenses against universal attacks}
\begin{figure*}[t]
\centering
\begin{subfigure}{.4\textwidth}
  \includegraphics[width=\linewidth]{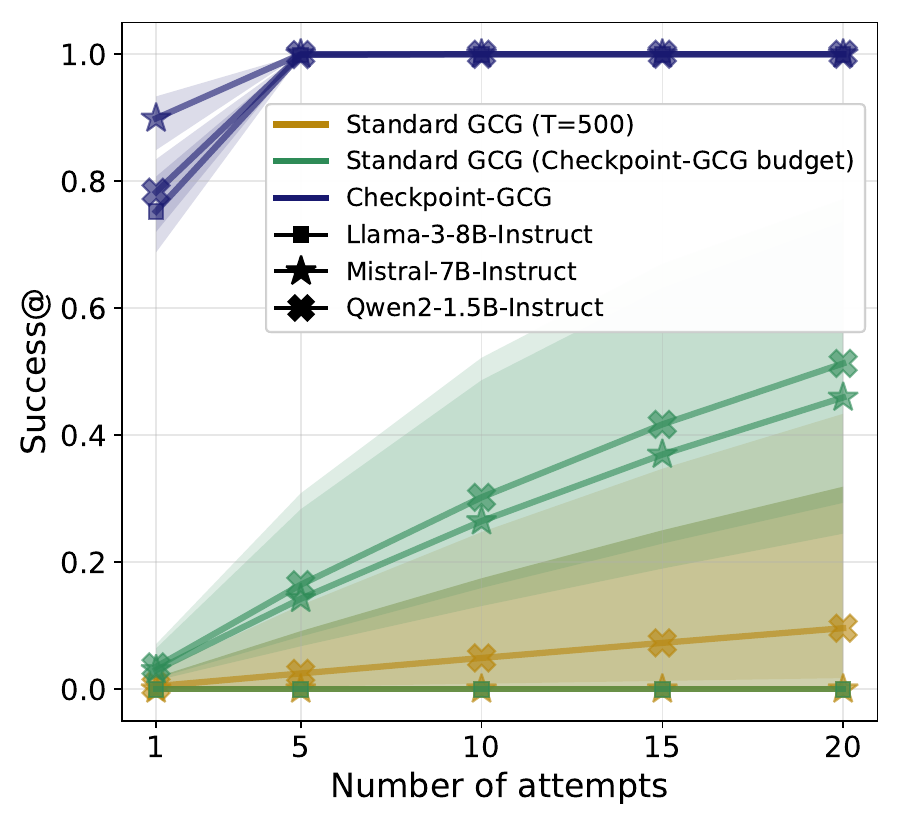}
  \caption{In-distribution}
  \label{fig:universal_plots_secalign_within_dataset}
\end{subfigure} %
\begin{subfigure}{.4\textwidth}
  \centering
  \includegraphics[width=\linewidth]{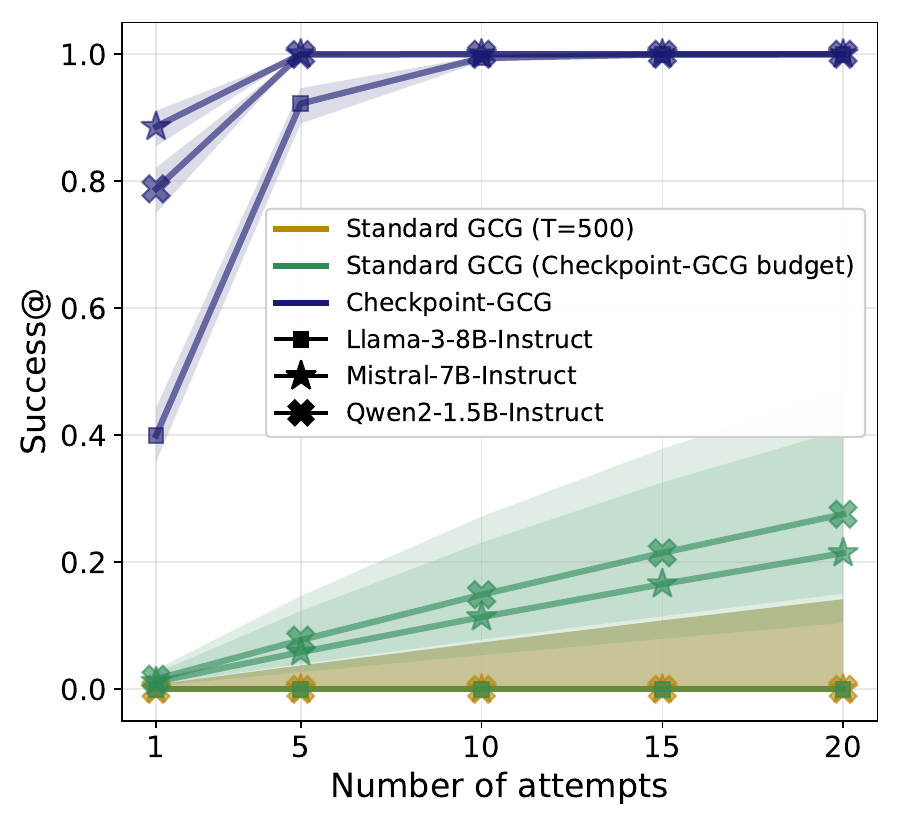}
  \caption{Out-of-distribution}
  \label{fig:universal_plots_secalign_cross_dataset}
\end{subfigure}
\caption{Universality of the Checkpoint-GCG suffixes on (a) in-distribution and (b) out-of-distribution test samples. Curves show \texttt{success@k} and shaded bands show 95\% confidence intervals obtained by propagating the Wilson score interval of the ASR through the \texttt{success@k} (see
 Sec.~\ref{sec:exp_setup}).
Results for SecAlign; for StruQ see App.~\ref{app:universal_results}.}
\label{fig:universal_plots_secalign}
\end{figure*}

Beyond prompt-specific attacks, we here instantiate Checkpoint-GCG to find a single \emph{universal} suffix that generalizes across prompts. Specifically, we optimize a suffix on $N_{\text{train}}=10$ training samples from AlpacaFarm. We then test the universality of the suffix (i.e. $s_{C, N_{\text{train}}}^{*}$) out-of-the-box against $\theta_C$ on the remaining $N_{\text{test}}=198$ held-out AlpacaFarm samples. To assess universality beyond the distribution of training samples, we also test on $N_{\text{test}}=500$ random samples from the SEP dataset~\citep{mu2023can}, reflecting a deployment-like scenario where attacks must generalize to unseen and potentially out-of-distribution inputs. Since \texttt{success@} is a monotonically increasing function of the ASR, we obtain confidence bands by computing the 95\% Wilson score interval for the ASR over the $N_{\text{test}}$ held-out samples and mapping its lower and upper bounds through the same \texttt{success@} expression; monotonicity guarantees the result is again a 95\%
interval.

Figure~\ref{fig:universal_plots_secalign_within_dataset} shows the universality (\texttt{success@k}) of attacks against SecAlign-defended models with in-distribution test samples (AlpacaFarm). We find that Checkpoint-GCG achieves a high probability of success even when restricted to a single attempt ($k=1$). For example, against Llama-3-8B-Instruct defended with SecAlign, it reaches \texttt{success@1} of $75.3\%$, 
while standard GCG is ineffective ($0\%$, with both 500 steps and Checkpoint-GCG budget). With just $k=5$ attempts, Checkpoint-GCG reaches almost perfect performance, whereas standard GCG maintains low \texttt{success@} values.
For test samples from SEP, Figure~\ref{fig:universal_plots_secalign_cross_dataset} shows that the absolute performance decreases, reflecting the greater difficulty of generalizing out-of-distribution. Nevertheless, Checkpoint-GCG still outperforms standard GCG by a wide margin, showing strong generalization both within and beyond the dataset used to construct the attack. 
More detailed results for the universal attack, including StruQ experiments showing the same overall pattern, are provided in App.~\ref{app:universal_results}.
\begin{table*}
\centering
\resizebox{\textwidth}{!}{ 
\begin{tabular}{ccccc}
    \toprule
    \multirow{2}{*}[-0.5em]{\parbox[b]{3.5cm}{\centering Universal suffix from $\theta_{C_s}$ obtained via}} & Black-box transfer to $\theta_{C_t}$ & \multicolumn{3}{c}{White-box transfer with Standard GCG on $\theta_{C_t}$} \\
    \cmidrule(lr){2-2} \cmidrule(lr){3-5}
    & ASR $\uparrow$ & Train ASR $\uparrow$ & Test ASR $\uparrow$ & $T$ steps $\downarrow$ \\
    \midrule
      Standard GCG (T=500) & 0 $[0, 1.9]$ & 0 $[0, 27.8]$ & 0 $[0, 1.9]$ & 5000   \\ 
      Standard GCG (Checkpoint-GCG budget) & 0 $[0, 1.9]$ & 0 $[0, 27.8]$ & 0 $[0, 1.9]$ & 5000   \\ 
      Checkpoint-GCG & $\mathbf{63.9}\ [57.2, 70.2]$ & $\mathbf{100}\ [72.2, 100]$ & $\mathbf{78.3}\ [72.0, 83.4]$ & $\mathbf{5}$  \\
    \bottomrule
\end{tabular}
}

\caption{
Attack success rate (ASR $\%$) for transferring the universal suffix found on the surrogate model (SecAlign-defended Llama-3-8B-Instruct) to the target model (Meta-SecAlign-8B, which is Llama-3.1-8B-Instruct defended with SecAlign++), in black- and white-box settings. The 95\% Wilson score intervals are shown in brackets.}

\label{tab:universal_suffix_transfer}
\end{table*}

\subsection{Towards realistic attacks}

\subsubsection{Relaxing assumption of intermediate checkpoints by interpolating model weights}\label{subsec:weaker_threat_model}
We extend the stepping-stones optimization approach to a weaker threat model and show that an attacker without access to intermediate fine-tuning checkpoints can achieve comparably effective attacks. Specifically, given an open-weights model, the attacker can construct a checkpoint trajectory by interpolating between the base (undefended) and final (defended) model weights. We demonstrate this by linearly interpolating between Llama-3-8B-Instruct and its SecAlign-defended counterpart, yielding 51 interpolated model states. Applying Checkpoint-GCG to these interpolated states achieves 90\% ASR on 20 samples (95\% Wilson score interval $[69.9, 97.2]$) with an average of 1,205 total steps, comparable to using actual fine-tuning checkpoints (Table~\ref{tab:checkpoint_strategy_results} in App.~\ref{app:checkpoint_strategy}).

Our auditing setup shows vulnerabilities exist under a strong threat model, while techniques like weight interpolation and transfer attacks (Sec.~\ref{subsec:realistic}) show these vulnerabilities are likely exploitable under weaker threat models and by future attacks. Together, we show that vulnerabilities discovered through auditing are not merely artifacts of the strong threat model but reflect genuine brittleness that realistic adversaries can exploit.

\subsubsection{Transferrable universal attacks}\label{subsec:realistic}
Beyond auditing, a strong threat model that leverages checkpoints can enable practical real-world attacks. An adversary can run Checkpoint-GCG on a checkpointed surrogate $\theta_{C_s}$ to obtain a universal suffix, then transfer it to a defended target $\theta_{C_t}$ with a different base model and defense and no accessible checkpoints. We evaluate black-box transfer (the attacker can only prompt the target) and white-box transfer (the attacker has the final target model's weights but no fine-tuning checkpoints, and can run additional optimization steps on the target). As target we use Meta-SecAlign-8B, a recently released model applying SecAlign++ to Llama-3.1-8B-Instruct.

Table~\ref{tab:universal_suffix_transfer} shows that Checkpoint-GCG suffixes transfer when the surrogate and target share similar models and defenses. Using SecAlign-defended Llama-3-8B-Instruct as surrogate, the suffix reaches $63.9\%$ ASR on Meta-SecAlign-8B black-box, and $78.3\%$ on 198 held-out samples after only 5 GCG optimization steps on 10 training samples. By contrast, a standard GCG suffix transfers at $0\%$ in both settings, even after 5{,}000 steps (500 per training sample). Transfer across substantially different model families remains limited (Mistral and Qwen surrogates yield $0\%$), but transfer within a family is an important real-world risk: SecAlign and SecAlign++ are openly available, and capable defended models are released on Hugging Face~\citep{metasecalignhf}, making them plausible targets for attacks optimized on a related model and transferred with minimal adaptation.

%% file: sections/6_discussion.tex
\textbf{Who is harmed, and who is responsible.} 
Defended models are an important part of the security boundary between users and untrusted content. When a defense reports near-zero attack success and ships, it becomes an implicit safety promise to users whose emails, documents, and credentials the model mediates, and to developers building on it. Our results show this promise can be overstated: using artifacts defenders already have raises attack success from near zero to 96\%, and the suffixes transfer to a separately released defended model (Sec.~\ref{subsec:realistic}). 

\textbf{Auditing as a pre-deployment responsibility.}
We argue that the developers should find these vulnerabilities before release, not users or attackers afterward.
Because fine-tuning produces checkpoints for free, our framework lets a developer stress-test a defense without inventing the next state-of-the-art attack, and at a computational cost comparable to standard evaluations (App.~\ref{subsec:comp_cost}).
When auditing surfaces a vulnerability, the same artifacts can inform mitigations, for example, folding discovered suffixes back into training (App.~\ref{app:defense_against_cgcg}).
We view this as a concrete, low-cost practice to shorten the feedback loop between deploying a defense and discovering that it fails.

App.~\ref{app:additional} reports further analyses, including attention analysis and a theoretical intuition for why checkpoints help, computational cost of Checkpoint-GCG, comparison to other GCG improvements, and candidate defenses.

%% file: sections/7_related_work.tex
\textbf{Improving optimization-based attacks.}
Research on optimization-based attacks has pursued several directions: improving efficiency, e.g., through better token selection and multi-token updates~\citep{li2025exploiting,li2024faster,liao2024amplegcg}; altering the optimization objective~\citep{wang2024attngcg,liu2024advancing}; and improving effectiveness against black-box targets~\citep{zou2023universal,sitawarin2024pal,hayase2024query,Labunets_2025}. Most relevant to us is the \emph{initialization}: several works observe that GCG's convergence and success are highly sensitive to it~\citep{jia2024improved,li2025exploiting,zhang2024enja,hayase2024query,wang2025functional}; for instance, initializing with a suffix that succeeded on a different sample improves both speed and success rates~\citep{jia2024improved}, and \citet{wang2025functional} demonstrate that deliberately misaligning safety-aligned chat models can help uncover successful jailbreak suffixes. Checkpoint-GCG addresses initialization by using intermediate fine-tuning checkpoints as stepping stones (Sec.~\ref{sec:results}).

\textbf{Prompt injection defenses.} 
Defenses against prompt injection attacks fall into two broad categories: system-level and model-level. System-level defenses include detection, often using a second LLM to identify injected instructions~\citep{liu2025datasentinel,inan2023llama}, prompt engineering~\citep{hines2024defending, benchmarkingpi}, and protective system layers around LLMs~\citep{debenedetti2025defeating}. Fine-tuning-based model-level defenses are an important part of the defense landscape and are the focus of this work. 

%% file: sections/8_conclusion.tex
Fine-tuning-based defenses against prompt injection~\citep{chen2024struq,chen2024aligning,wallace2024instruction,wu2024instructional} are typically evaluated against the best realistic attack available at release, which we argue is insufficient. Our key insight is that the fine-tuning process itself yields intermediate checkpoints, that, although unavailable to real attackers, expose genuine vulnerabilities in the final defended model. We introduce a strong threat model and an attack-agnostic auditing framework that uses checkpoints as stepping stones; instantiated with GCG, it raises attack success from near zero to up to 96\% against SecAlign across three models, and instantiating with BEAST~\citep{sadasivan2024fastadversarialattackslanguage} yields similar gains. These vulnerabilities are exploitable by realistic adversaries: universal suffixes transfer across models sharing a related defense, and an attacker with access only to the base and defended models can reconstruct a checkpoint trajectory by interpolation.

Robustness evaluations should therefore go beyond the best realistic attack of the day and explicitly include strong pre-deployment audits. Because the required artifacts already exist during training, developers can run this audit cheaply, identifying and closing these vulnerabilities before the defenses, and the products built on them, reach users.

%% file: sections/9_ethical.tex
Our work affects several stakeholders. Developers and providers of LLMs and defenses rely on empirical evaluations to deploy secure models; our results may reveal weaknesses in existing defenses, informing better design while also exposing potential misuse. End users, whose interactions with models may be sensitive, are affected by the vulnerabilities we uncover, and publishing our results raises awareness of these risks and encourages safer deployment. More broadly, society and the research community benefit from reliable safety evaluation: by exposing subtle vulnerabilities and proposing systematic auditing strategies, we provide methods that foster reproducibility, collaboration, and stronger defenses across the field.

Our work demonstrates that existing empirical evaluations may miss adversarial vulnerabilities. By proposing a strong threat model that leverages intermediate checkpoints for progressive attack optimization, we provide a more thorough robustness analysis, showing that defenses appearing robust under realistic attacks may still be vulnerable. This contributes positively by informing safer deployment, but also highlights risks if attackers were to leverage similar methods.

To mitigate these risks, all experiments were conducted on publicly available or open-source models, defenses, and datasets, and our methodological descriptions focus on the process of constructing attacks to evaluate robustness rather than on concrete adversarial instructions that could be directly exploited. We further emphasize responsible disclosure and frame our work as a systematic auditing approach rather than an attack manual. We chose to conduct and publish this research because prompt injection is a critical threat to deployed LLMs, which often handle sensitive data in safety-critical contexts; the strong attacker framework we study provides auditing tools that help identify vulnerabilities before they can be exploited, and it carries a low risk of misuse because the assumed threat model is not easily accessible to anyone other than model developers.

%% file: appendix/appendix.tex
\input{appendix/b_checkpoint_selection_strategies}

\input{appendix/other_attacks}
\input{appendix/a_detailed_results}

\input{appendix/c_early_stopping}
\input{appendix/additional_analyses}

\input{appendix/h_replicate_struq_secalign}

\input{appendix/f_jailbreak}
\input{appendix/g_finetuning_process}

\input{appendix/i_computational_resources}

%% file: appendix/b_checkpoint_selection_strategies.tex
\section{Checkpoint selection}
\label{app:checkpoint_strategy}

An auditor could instantiate our framework using all checkpoints available. However, choosing a subset of checkpoints can improve computational efficiency while maintaining attack effectiveness. 

We consider four strategies for selecting a subset $\mathcal{S} = [c_1, \dots, c_k] \subseteq \mathcal{I}$ of checkpoint indices to attack. 
All strategies include the base model ($c_1 = 0$) and final checkpoint ($c_k = C$), and distinctly select intermediate checkpoints ($0 < c_i < C$):

\textit{1. Frequency-based (\textsc{freq}).} For simplicity and to provide uniform coverage of the training process, we select every $q^{th}$ checkpoint, i.e., $\mathcal{S}_\textsc{freq} = \{c \in \mathcal{I} \mid c = q\cdot l,\; l \in \mathbb{N}_0\}$.

\textit{2. Step-based (\textsc{step}).} Since the most substantial changes to model parameters typically occur in the early stages of training, we select all checkpoints up to a training step $r$ to capture these changes. To maintain coverage throughout training, we also include every $q^{th}$ checkpoint thereafter, i.e., $\mathcal{S}_\textsc{step} = \{c \in \mathcal{I} \mid c \leq r \} \cup \{c \in \mathcal{I} \mid c> r, c = q\cdot l,\; l \in \mathbb{N}_0\}$.

\textit{3. Loss-based (\textsc{loss}).} As training loss $\mathcal{L}_{\theta_c}$ represents the model error and guides the updates of model parameters, we select a checkpoint if its training loss differs from the loss at the last selected checkpoint by at least a threshold $\tau_\text{loss}$, i.e., $\mathcal{S}_{\textsc{loss}} = \{ c \in \mathcal{I} \mid\ |\mathcal{L}_{\theta_{c}} - \mathcal{L}_{\theta_{s}}| \geq \tau_{\text{loss}},s=max\{x\in\mathcal{S_{\textsc{loss}}}|x<c\} \}$. Additionally, to ensure coverage during periods of low change, we include every $q^{\text{th}}$ checkpoint when this condition is not met for $q$ consecutive steps. 

\textit{4. Gradient-based (\textsc{grad}).}  Gradient norms $\|\nabla_{\theta_c} \mathcal{L}_{\theta_c} \|$ provide a more direct measure of the magnitude of updates made to model parameters at every step. We therefore select checkpoints where the gradient norm is at least a threshold $\tau_\text{grad}$, indicating that the model is making sufficient changes, i.e., $\mathcal{S}_{\textsc{grad}} = \{ c \in \mathcal{I} \mid\ \|\nabla_{\theta_c} \mathcal{L}_{\theta_c} \| \geq \tau_{\text{grad}} \}$.

\begin{figure*}[h]
    \centering
    \begin{subfigure}{.32\textwidth}
        \centering
        \includegraphics[width=\linewidth]{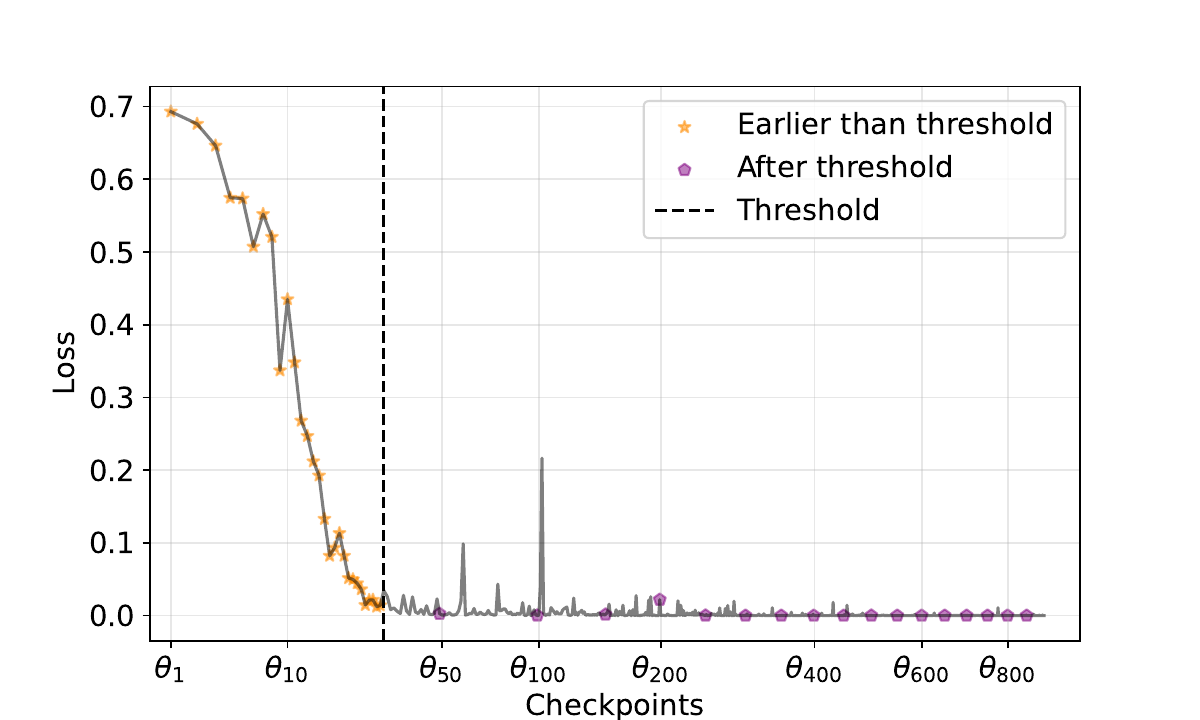}
        \caption{\textsc{step}, $r=30\; \&\; q=50$}
        \label{fig:selectedcheckpointsstep}
    \end{subfigure}
    \begin{subfigure}{.32\textwidth}
        \centering
        \includegraphics[width=\linewidth]{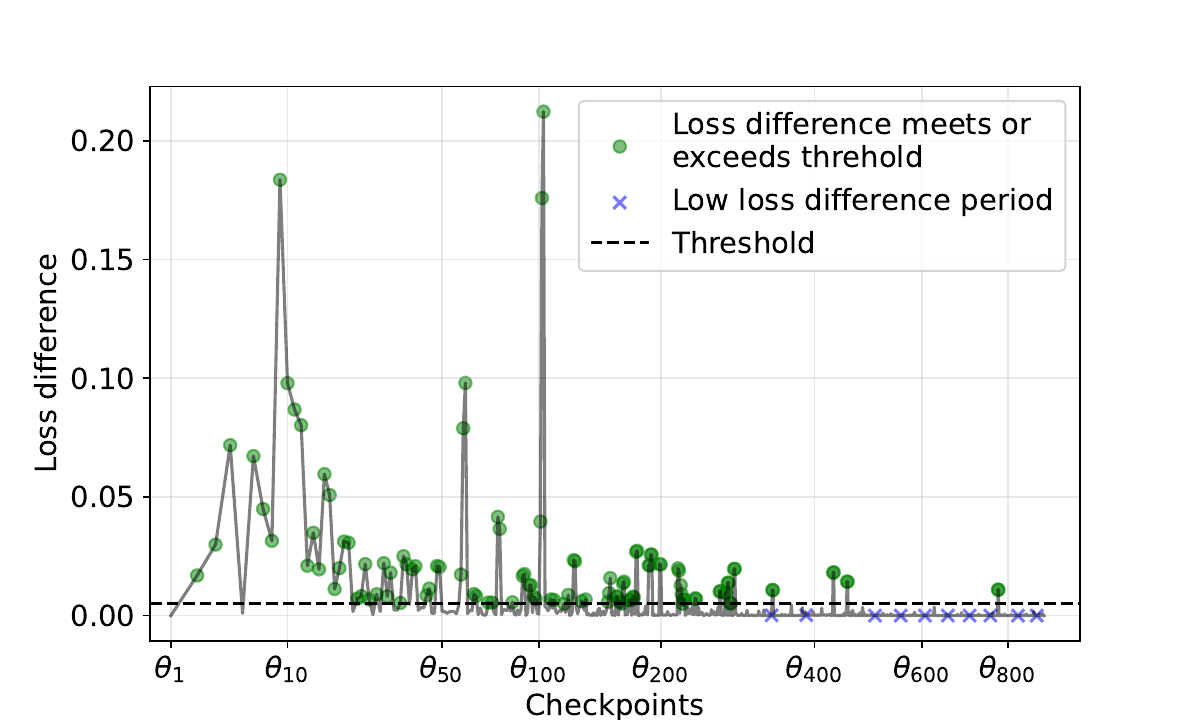}
        \caption{\textsc{loss}, $\tau_\text{loss}=0.005\; \&\; q=50$}
        \label{fig:selectedcheckpointsloss}
    \end{subfigure}
    \begin{subfigure}{.32\textwidth}
        \centering
        \includegraphics[width=\linewidth]{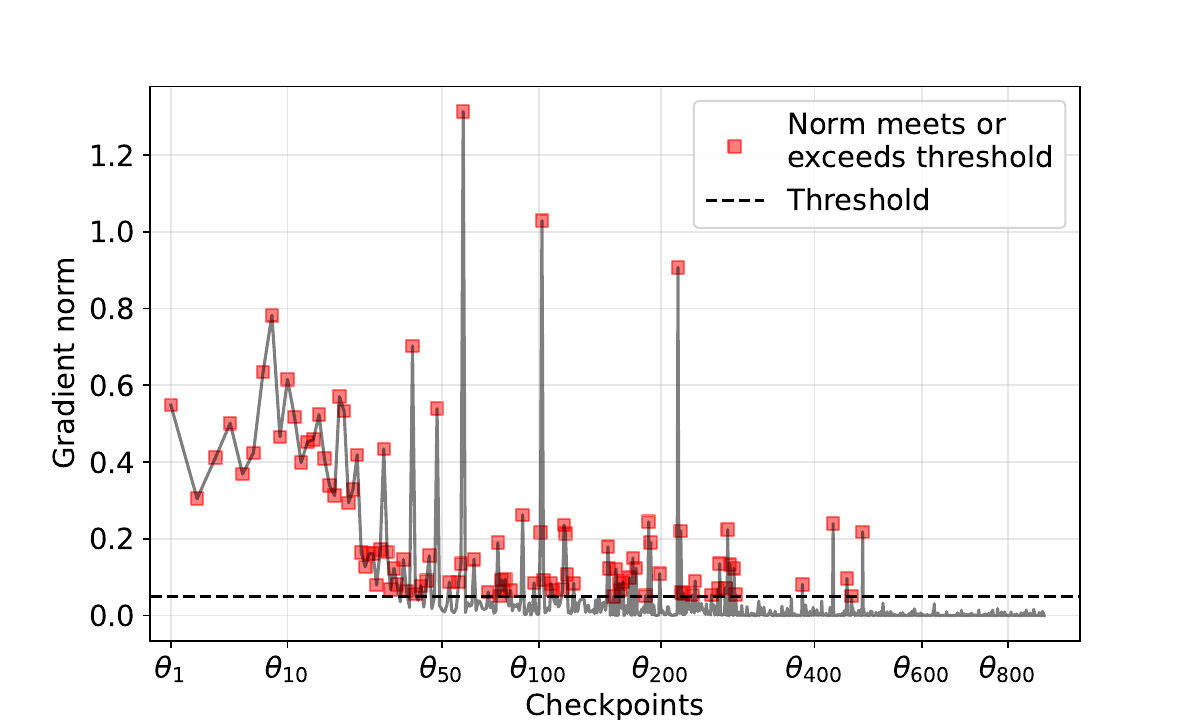}
        \caption{\textsc{grad}, $\tau_\text{grad}=0.05$}
        \label{fig:selectedcheckpointsgradnorm}
    \end{subfigure}
    
    \caption{Illustrating the fine-tuning checkpoints selected for SecAlign-defended Llama-3-8B-Instruct, using \textsc{step}, \textsc{loss} and \textsc{grad} with an example hyperparameter configuration for each strategy (see App.~\ref{app:checkpoint_strategy}).}
    
    \label{fig:selectedcheckpoints}
\end{figure*}

Figure~\ref{fig:selectedcheckpoints} visualizes the checkpoints selected by \textsc{step}, \textsc{loss} and \textsc{grad} strategies from performing SecAlign fine-tuning on Llama-3-8B-Instruct.

We assess both the empirical effectiveness and the computational efficiency of the four checkpoint selection strategies defined above. 
Using SecAlign-defended Llama-3-8B-Instruct as a representative model, we vary the hyperparameters for each strategy and summarize the corresponding results in Table~\ref{tab:checkpoint_strategy_results}.

\begin{table*}[h]
\centering
\resizebox{0.9\textwidth}{!}{%
\begin{tabular}{cc|ccc}
    \toprule
    Checkpoint  & \multirow{2}{*}{Parameter values} &  \multirow{2}{*}{ASR (\%) $\uparrow$} & \# Selected & Total Checkpoint-GCG steps \\
    strategy &  &  & checkpoints $\downarrow$ &  (avg across samples) $\downarrow$ \\
    \midrule
    All & / & \multirow{1}{*}{\textbf{100}} & 898 & \multirow{1}{*}{4,228}  \\
    \midrule
    \multirow{3}{*}{\textsc{freq}} 
    & $q=10$ & 95 & 91 & 4,037 \\    
     & $q=50$ & 65 & 19 & 4,676 \\    
     & $q=100$ & 65 & 10 & 2,659 \\   \midrule 
    \multirow{3}{*}{\textsc{step}} 
    & $r=30\; \&\; q=10$ & \textbf{100} & 118 & 3,708 \\ 
     & $r=30\; \&\; q=50$ & 75 & 49 & 2,873 \\
     & $r=30\; \&\; q=100$ & 85 & 40 & 1,553 \\   
    \midrule 
    \textsc{loss} 
    & $\tau_\text{loss}=0.005\; \&\; q=50$ & \textbf{100} & 124 & 3,754 \\
    \midrule 
    \multirow{7}{*}{\textsc{grad}}
    & $\tau_\text{grad}=0.05$ & \textbf{100} & 102 & 3,077 \\
    & $\tau_\text{grad}=0.1$ & \textbf{100} & 64 & 2,033 \\
    & $\tau_\text{grad}=0.2$ & 90 & 41 & 1,764 \\
    & $\tau_\text{grad}=0.3$ & 90 & 29 & 1,475 \\
    & $\tau_\text{grad}=0.4$ & 50 & 19 & 1,721 \\
    & $\tau_\text{grad}=0.5$ & 45 & 12 & 1,213 \\
    & $\tau_\text{grad}=0.6$ & 20 & \textbf{5} & \textbf{798} \\
    \bottomrule
\end{tabular}%
}

\caption{Attack effectiveness (ASR) and computational cost (number of selected checkpoints, and total Checkpoint-GCG steps averaged across samples) for each checkpoint selection strategy, evaluated on SecAlign-defended~\citep{chen2024aligning} Llama-3-8B-Instruct~\citep{llama3modelcard}. Results are reported over $20$ randomly selected samples from AlpacaFarm~\citep{dubois2023alpacafarm}. Each Checkpoint-GCG step takes roughly constant time (about 3s on our hardware: A100 GPUs with 80GB RAM), identical to standard GCG.}

\label{tab:checkpoint_strategy_results}
\end{table*}

These results show that checkpoint selection requires balancing attack performance and computational efficiency. The gradient-based strategy (\textsc{grad}) offers the best balance between attack effectiveness (ASR) and computational cost (Total Checkpoint-GCG steps). While \textsc{loss} and \textsc{step} also achieve the same ASR with some of the hyperparameter values, they require a higher computational cost. 
The \textsc{freq} strategy, on the other hand, struggles to achieve the same ASR even with a higher computational cost.

\begin{figure}[t]
    \centering
        \centering
        \includegraphics[width=0.6\linewidth]{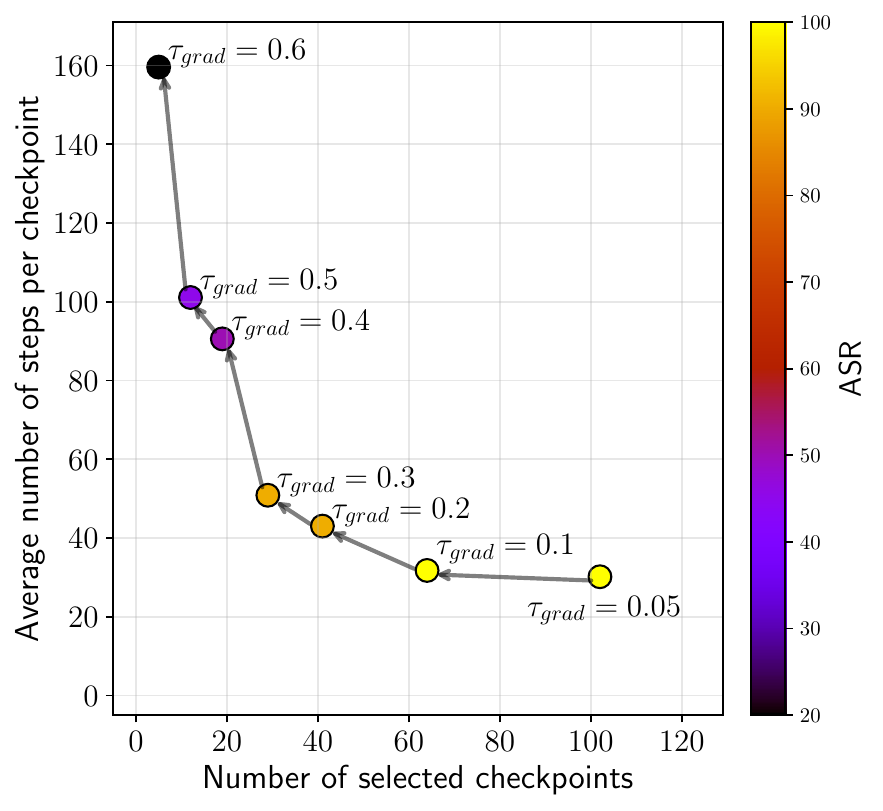}
    \caption{Trade-off between number of selected checkpoints, average number of steps per checkpoint, and the Checkpoint-GCG ASR for the \textsc{grad} checkpoint selection strategy.}
        \label{fig:checkpoint_selection_strategies}
\end{figure}

Figure~\ref{fig:checkpoint_selection_strategies} illustrates the trade-off between the number of selected checkpoints, the average number of GCG steps required at each checkpoint, and the resulting ASR. When too few checkpoints are selected, Checkpoint-GCG lacks sufficient stepping stones, leading to lower ASR. As the number of selected checkpoints increases, the ASR improves and, notably, the number of GCG steps required per checkpoint decreases. This is expected: selecting more checkpoints leads to smaller parameter changes between them, which reduces the adjustments needed to update the adversarial suffixes. Beyond a certain point, adding more checkpoints may however increase the cumulative number of steps across all checkpoints without yielding proportional gains in ASR. 

Selecting appropriate \textsc{grad} thresholds helps strike an effective balance: by choosing checkpoints with significant parameter updates, it ensures that each attack always starts from a well-informed initialization and targets a meaningful transition in the model's behavior. We therefore adopt the \textsc{grad} strategy for all experiments reported in Section~\ref{sec:results}. 

For Llama-3-8B-Instruct defended with SecAlign, we choose a threshold of $\tau_\text{grad}=0.05$, although the computational cost may be further reduced by choosing a higher threshold, as shown in Table~\ref{tab:checkpoint_strategy_results}. 
For all other models and defenses, we choose the values for $\tau_{\text{grad}}$ such that a similar number of checkpoints are selected. Table~\ref{tabapp:checkpoint_strategy_values} shows the values of $\tau_{\text{grad}}$ for different defenses and models and the resulting number of selected checkpoints.

\begin{table}[h]
\centering
\begin{tabular}{cccc}
\toprule
Defense & Model & $\tau_\text{grad}$ & \# Checkpoints \\
\midrule
\multirow{3}{*}{\parbox{2.5cm}{\centering StruQ~\citep{chen2024struq}}} 
& Llama-3-8B-Instruct~\citep{llama3modelcard} & 4.5 & 127 \\
& Mistral-7B-Instruct~\citep{jiang2023mistral} & 7 & 112 \\
& Qwen2-1.5B-Instruct~\citep{qwen2} & 3.2 & 100 \\
\midrule
\multirow{3}{*}{\parbox{1.5cm}{\centering SecAlign~\citep{chen2024aligning}}} 
& Llama-3-8B-Instruct~\citep{llama3modelcard} & 0.05 & 102 \\
& Mistral-7B-Instruct~\citep{jiang2023mistral} &  0.05 & 93 \\
& Qwen2-1.5B-Instruct~\citep{qwen2} & 0.12 & 91 \\
\bottomrule
\end{tabular}
\caption{Parameters for the \textsc{grad} checkpoint selection strategy used across experiments. We provide both the selected gradient norm threshold $\tau_\text{grad}$ and the resulting number of checkpoints selected using this threshold.}
\label{tabapp:checkpoint_strategy_values}
\end{table}

%% file: appendix/other_attacks.tex
\section{Instantiating the framework with other attack algorithms}\label{subsec:other_attack_algos}
We focus on GCG as the seminal optimization-based attack and primary baseline used by defenses to demonstrate robustness. 
Our progressive optimization framework is, however, algorithm-agnostic and can be instantiated with other optimization-based attacks. To show this, we instantiate it with BEAST~\citep{sadasivan2024fastadversarialattackslanguage}, a gradient-free beam search method that allows the attacker to balance between attack speed, success rate, and the readability of adversarial prompts by tuning attack parameters. Since BEAST appends suffix tokens autoregressively, we adapt it to our warm-starting framework by fixing the suffix length: at each step, each beam candidate generates $k_2$ neighbors by resampling a random suffix position. On 20 AlpacaFarm samples against SecAlign-defended Llama-3-8B-Instruct, Checkpoint-BEAST achieves 80\% ASR, whereas BEAST attacking the final model directly achieves 0\%, confirming that our framework benefits attacks with different optimization dynamics. 

%% file: appendix/a_detailed_results.tex
\section{Detailed results for Checkpoint-GCG}\label{app:mainresults}

\subsection{Auditing defenses}
We present the fine-grained ASRs obtained by each method against each of the defenses. 
Evaluations based solely on GCG, both with $T=500$ steps and as many steps as Checkpoint-GCG (i.e., Checkpoint-GCG budget), suggest that robustness increases substantially with more advanced defenses.
However, when audited under the stronger checkpoint-based threat model, these same defenses exhibit significant vulnerabilities. For example, standard GCG struggles the most against Llama-3-8B-Instruct~\citep{llama3modelcard} protected by the state-of-the-art defense SecAlign~\citep{chen2024aligning}, achieving up to 12\% ASR, while Checkpoint-GCG achieves an ASR of 88\%.

\begin{table*}[h]
\centering
\resizebox{\textwidth}{!}{ 
\begin{tabular}{ccccc}
    \toprule
     & & \multicolumn{2}{c}{Standard GCG on $\theta_C$} & \\
    \cmidrule(lr){3-4}
    \multirow{1}{*}{Defense} & \multirow{1}{*}{Model} &  $T=500$ steps & Checkpoint-GCG budget & Checkpoint-GCG (ours) \\
    \midrule
    \multirow{3}{*}{\parbox{2cm}{\centering Undefended}} & Llama-3-8B-Instruct~\citep{llama3modelcard} & \textbf{100} & \textbf{100} & \textbf{100} \\ 
     & Mistral-7B-Instruct~\citep{jiang2023mistral}  & \textbf{100} & \textbf{100} & \textbf{100} \\
     & Qwen2-1.5B-Instruct~\citep{qwen2} & \textbf{98} & \textbf{98} & \textbf{98} \\
    \midrule
    \multirow{3}{*}{\parbox{2cm}{\centering StruQ~\citep{chen2024struq}}} & Llama-3-8B-Instruct~\citep{llama3modelcard} & 26 & 50 & \textbf{100} \\ 
     & Mistral-7B-Instruct~\citep{jiang2023mistral}  & 88 & 90 & \textbf{100} \\
     & Qwen2-1.5B-Instruct~\citep{qwen2} & 48 & 52 & \textbf{90} \\
    \midrule
    \multirow{3}{*}{\parbox{2cm}{\centering SecAlign~\citep{chen2024aligning}}} & Llama-3-8B-Instruct~\citep{llama3modelcard} & 6 & 12 & \textbf{88} \\ 
     & Mistral-7B-Instruct~\citep{jiang2023mistral} & 18 & 22 & \textbf{96} \\
     & Qwen2-1.5B-Instruct~\citep{qwen2} & 32 & 32 & \textbf{84} \\
    \bottomrule
\end{tabular}%
}
\caption{Attack success rate (ASR $\%$) $\uparrow$ for Checkpoint-GCG against state-of-the-art prompt injection defenses. As baseline, we apply the standard GCG attack directly to the defended model (i.e., the final checkpoint $\theta_C$). Results are aggregated for $50$ randomly selected samples from AlpacaFarm~\citep{dubois2023alpacafarm}.}
\label{tab:mainresults}
\end{table*}

\subsection{Universal attack}
\label{app:universal_results}
\begin{table*}[h]
\centering
\resizebox{\textwidth}{!}{ 
\begin{tabular}{cccccccc}
    \toprule
     \multirow{2}{*}{Defense} & \multirow{2}{*}{Model} & \multicolumn{2}{c}{Standard GCG with 500 steps per sample} & \multicolumn{2}{c}{Standard GCG with Checkpoint-GCG budget} & \multicolumn{2}{c}{Checkpoint-GCG (ours)} \\
    \cmidrule(lr){3-4} \cmidrule(lr){5-6} \cmidrule(lr){7-8}
     &  & Training & Testing & Training & Testing & Training & Testing \\
    \midrule
    \multirow{3}{*}{\parbox{2cm}{\centering SecAlign~\citep{chen2024aligning}}} 
        & Llama~\citep{llama3modelcard} & 0 & 0 & 0 & 0 & \textbf{100} & \textbf{75.3} \\ 
        & Mistral~\citep{jiang2023mistral} & 0 & 0 & 0 & 3.0 & \textbf{100} & \textbf{89.9} \\
        & Qwen~\citep{qwen2} & 0 & 0.5 & 0 & 3.5 & \textbf{100} & \textbf{78.3} \\
    \midrule
    \multirow{3}{*}{\parbox{2cm}{\centering StruQ~\citep{chen2024struq}}} 
        & Llama~\citep{llama3modelcard} & 0 & 0 & 70 & 74.2 & \textbf{100} & \textbf{88.9} \\ 
        & Mistral~\citep{jiang2023mistral} & 30 & 58.1 & \textbf{100} & 91.4 & \textbf{100} & \textbf{99.0} \\
        & Qwen~\citep{qwen2} & 10 & 2.0 & 40 & 27.8 & \textbf{100} & \textbf{87.9} \\
    \bottomrule
\end{tabular}%
}
\caption{Attack success rate (ASR $\%$) $\uparrow$ for \textbf{in-distribution universal attack} comparing standard GCG with 500 steps per training sample, standard GCG with Checkpoint-GCG budget, and our Checkpoint-GCG method across defenses and models. Results are reported on a training set of 10 samples and a testing set of 198 samples from AlpacaFarm~\citep{dubois2023alpacafarm}.}
\label{tab:universal_main_results}
\end{table*}

\begin{table*}[h]
\centering
\resizebox{\textwidth}{!}{ 
\begin{tabular}{ccccc}
    \toprule
     \multirow{1}{*}{Defense} & \multirow{1}{*}{Model} & \multicolumn{1}{c}{Standard GCG with 500 steps per sample} & \multicolumn{1}{c}{Standard GCG with Checkpoint-GCG budget} & \multicolumn{1}{c}{Checkpoint-GCG (ours)} \\
    \midrule
    \multirow{3}{*}{\parbox{2cm}{\centering SecAlign~\citep{chen2024aligning}}} 
        & Llama~\citep{llama3modelcard} & 0 & 0 & \textbf{40.0} \\ 
        & Mistral~\citep{jiang2023mistral} & 0 & 0.01 & \textbf{88.6} \\
        & Qwen~\citep{qwen2} & 0 & 0.01 & \textbf{78.8} \\
    \midrule
    \multirow{3}{*}{\parbox{2cm}{\centering StruQ~\citep{chen2024struq}}} 
        & Llama~\citep{llama3modelcard} & 0 & 83.8 & \textbf{96.2} \\ 
        & Mistral~\citep{jiang2023mistral} & 65.0 & 86.2 & \textbf{95.8} \\
        & Qwen~\citep{qwen2} & 0.03 & 40.6 & \textbf{91.6} \\
    \bottomrule
\end{tabular}%
}
\caption{Attack success rate (ASR $\%$) $\uparrow$ for \textbf{out-of-distribution universal attack} comparing standard GCG with 500 steps per training sample, standard GCG with Checkpoint-GCG budget, and our Checkpoint-GCG method across defenses and models. Results are reported on a testing set of 500 samples from SEP~\citep{mu2023can}.}
\label{tab:universal_main_results_out_distribution}
\end{table*}

Tables~\ref{tab:universal_main_results} and \ref{tab:universal_main_results_out_distribution} show the detailed ASRs achieved by each attack per model and defense for in-distribution and out-of-distribution samples, respectively. While standard GCG struggles to find a universal suffix against the stronger SecAlign defense that works across the training samples, Checkpoint-GCG finds suffixes that are successful on all 10 training samples and also generalize to other unseen samples. Even on the weaker StruQ defense, Checkpoint-GCG consistently finds universal suffixes that generalize better than the suffixes discovered by standard GCG.

\begin{figure}
\centering
\begin{subfigure}{.3\textwidth}
  \centering
  \includegraphics[width=\linewidth]{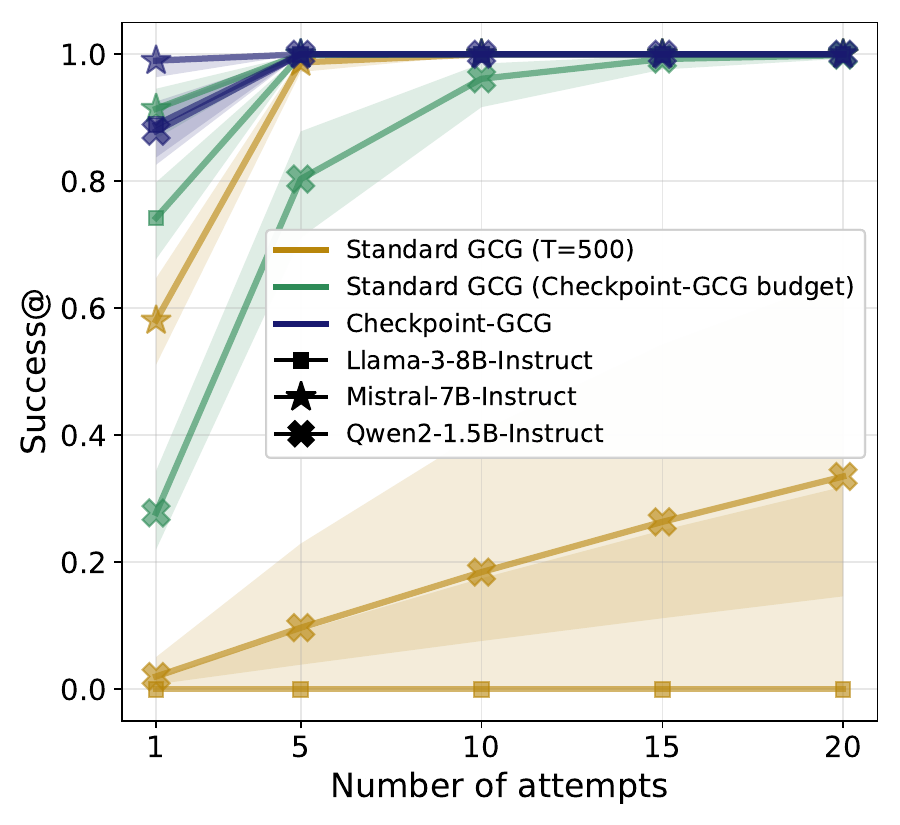}
  \caption{In-distribution}
  \label{appfig:universal_plots_struq_within_dataset}
\end{subfigure} %
\begin{subfigure}{.3\textwidth}
  \centering
  \includegraphics[width=\linewidth]{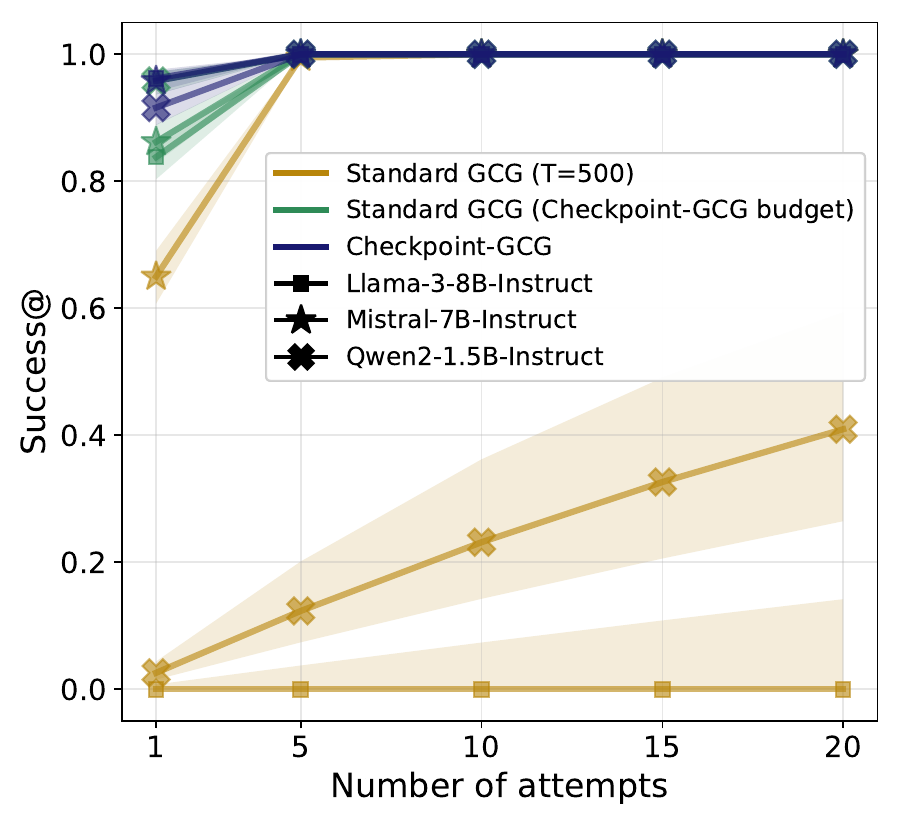}
  \caption{Out-of-distribution}
  \label{appfig:universal_plots_struq_cross_dataset}
\end{subfigure}
\caption{Universality of suffixes discovered by Checkpoint-GCG on (a) in-distribution and (b) out-of-distribution test samples. Curves show \texttt{success@k} and shaded bands show 95\% confidence intervals obtained by propagating the Wilson score interval of the ASR through the \texttt{success@k} (see Sec.~\ref{sec:exp_setup}). Results shown are for suffixes discovered against StruQ.}
\label{appfig:universal_plots_struq}
\end{figure}

Figure~\ref{appfig:universal_plots_struq} shows the universality of suffixes discovered by Checkpoint-GCG when evaluated on unseen samples from two datasets against StruQ-defended models. We note that Mistral-7B-Instruct, defended with StruQ, is noticeably less robust than the others. However, Checkpoint-GCG still consistently outperforms standard GCG (both with 500 steps and with Checkpoint-GCG budget).

\subsubsection{Ablation on number of tokens for universal suffix}
We ablate the number of suffix tokens for universal suffix by instantiating Checkpoint-GCG against SecAlign-defended Llama-3-8B-Instruct. 
We increase the suffix length from 20 to 25 and 30 tokens, and find that the ASR on held-out AlpacaFarm test samples drops to 64.1\% and 45.5\%, respectively (Table~\ref{tab:suffix_length_asr}). This suggests that longer suffixes may overfit to the training samples. We leave for future work how to balance training and testing performance in universal suffix generation.

\begin{table}
\centering
\resizebox{0.6\textwidth}{!}{%
\begin{tabular}{c|cc}
    \toprule
    Suffix length (\# tokens) & Training ASR $\uparrow$ & Testing ASR $\uparrow$ \\
    \midrule
    20 & 100\% & 75.3\% \\
    25 & 100\% & 64.1\% \\
    30 & 100\% & 45.5\% \\
    \bottomrule
\end{tabular}%
}
\caption{Attack success rate (ASR $\%$) $\uparrow$ on $N_{\text{train}}=10$ training and the remaining $N_{\text{test}}=198$ held-out testing samples from AlpacaFarm across different suffix lengths.}
\label{tab:suffix_length_asr}
\end{table}

%% file: appendix/c_early_stopping.tex
\section{Early stopping}\label{app:early_stopping} 
In the original GCG algorithm, GCG terminates either when a successful suffix is found or after a fixed budget of $T=500$ steps. 
Since we are targeting models that have been specifically fine-tuned to be robust against attacks, we anticipate the attack to be more challenging and hence consider a per-checkpoint budget of $T=1,000$. 
To avoid excessive computation, we also implement \emph{early stopping}. Our observations show that GCG can get stuck in local minima, where it continues to iterate without improving the loss or finding a successful suffix. To mitigate this, Checkpoint-GCG terminates for checkpoint $\theta_{c_i}$ if the best GCG loss achieved for $\theta_{c_i}$ remains essentially unchanged (change $\leq1\mathrm{e}{-5}$) over 250 consecutive steps. These thresholds were selected empirically and on the conservative side, so that it's unlikely for Checkpoint-GCG to miss successful suffixes due to early-stopping. If early-stopping occurs at checkpoint $\theta_{c_i}$, Checkpoint-GCG proceeds to attacking the next checkpoint $\theta_{c_{i+1}}$, using the best suffix (i.e., the one with lowest loss) found at $\theta_{c_i}$ as initialization (see Algorithm~\ref{alg:checkpoint-gcg}). 

%% file: appendix/additional_analyses.tex
\section{Additional analyses}\label{app:additional}

\subsection{Checkpoints steer the optimization in the right direction}
\begin{figure*}[t]
    \centering
    \includegraphics[width=0.9\linewidth]{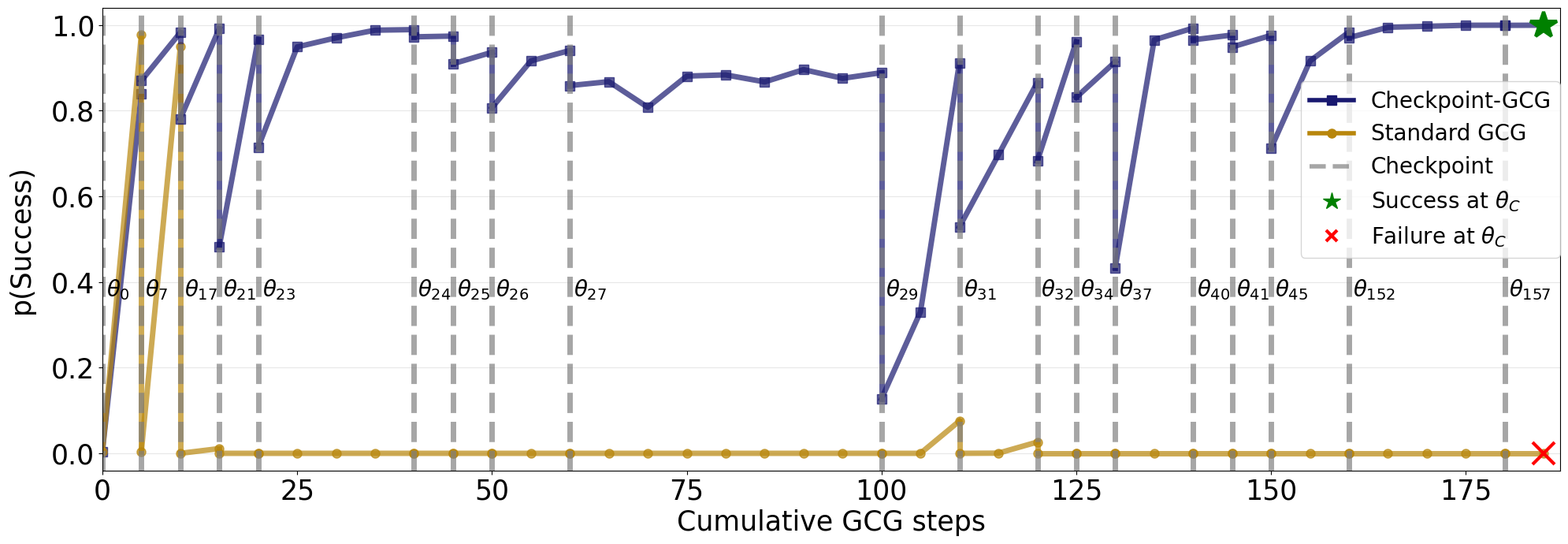} 
    \caption{The probability of a successful attack by GCG and Checkpoint-GCG when attacking one AlpacaFarm sample on SecAlign-defended~\citep{chen2024aligning} Llama-3-8B-Instruct~\citep{llama3modelcard}, over the cumulative number of optimization steps across checkpoints attacked. \texttt{p(Success)} is the probability of the model outputting the target sequence $y^*$, calculated given the cross entropy loss.}
    \label{fig:ptarget_vs_step}
\end{figure*}

We apply Checkpoint-GCG to find an adversarial suffix for a prompt injection attack against Llama-3-8B-Instruct~\citep{llama3modelcard} defended SecAlign~\citep{chen2024aligning}. 
Figure~\ref{fig:ptarget_vs_step} visualizes the optimization for one sample, showing the probability of attack success over the cumulative number of optimization steps across checkpoints. Any dashed vertical line denotes a checkpoint $\theta_c$ selected to attack.

We start by applying GCG on the base model parameters $\theta_0$, initializing the attack as in prior work \texttt{"!!!"}. 
For this suffix $s^{(0)}_{c=0}$, the probability of attack success is near $0$ (lower left of Figure~\ref{fig:ptarget_vs_step}). 
After a limited number of optimization steps, we find a suffix $s^{*}_{c=0}$ that successfully attacks the base model $\theta_0$. 
We then attack the next checkpoint, $\theta_7$, initializing GCG the successful suffix found against $\theta_0$. We find the success probability of this suffix to remain highly similar for $\theta_7$, yet a few optimization steps are needed to update $s^{*}_{c=0} = s^{(0)}_{c=7}$ to $s^{*}_{c=7}$, which successfully attacks $\theta_7$. We continue this process across all selected checkpoints. 
While the probability of success often drops going from checkpoints $\theta_{c_i}$ to $\theta_{c_{i+1}}$, applying a limited number of optimization steps starting from the suffix successful for $\theta_{c_i}$ quickly restores the success probability against $\theta_{c_{i+1}}$. Finally, Checkpoint-GCG applies the same strategy to the fully aligned model $\theta_C$, and finds the optimized suffix to succeed. 

As a reference, we also report the results for standard GCG when applied independently on each checkpoint $\theta_{c_i}$. At each $\theta_{c_i}$, we run standard GCG for the same number of steps as Checkpoint-GCG, but initialize the naive suffix (\texttt{"!!!"}) rather than the suffix obtained from optimizing against $\theta_{c_{i-1}}$. While standard GCG still improves success probability at early checkpoints, the fine-tuning process increasingly suppresses the attack at later stages. After only a few fine-tuning checkpoints, the success probability plateaus near zero, ultimately resulting in a failed attack on $\theta_C$.

\subsection{Understanding the superior performance through attention analysis}\label{subsec:interp}

\begin{figure*}[t]
\centering
\begin{subfigure}{.3\textwidth}
  \centering
  \includegraphics[width=\linewidth]{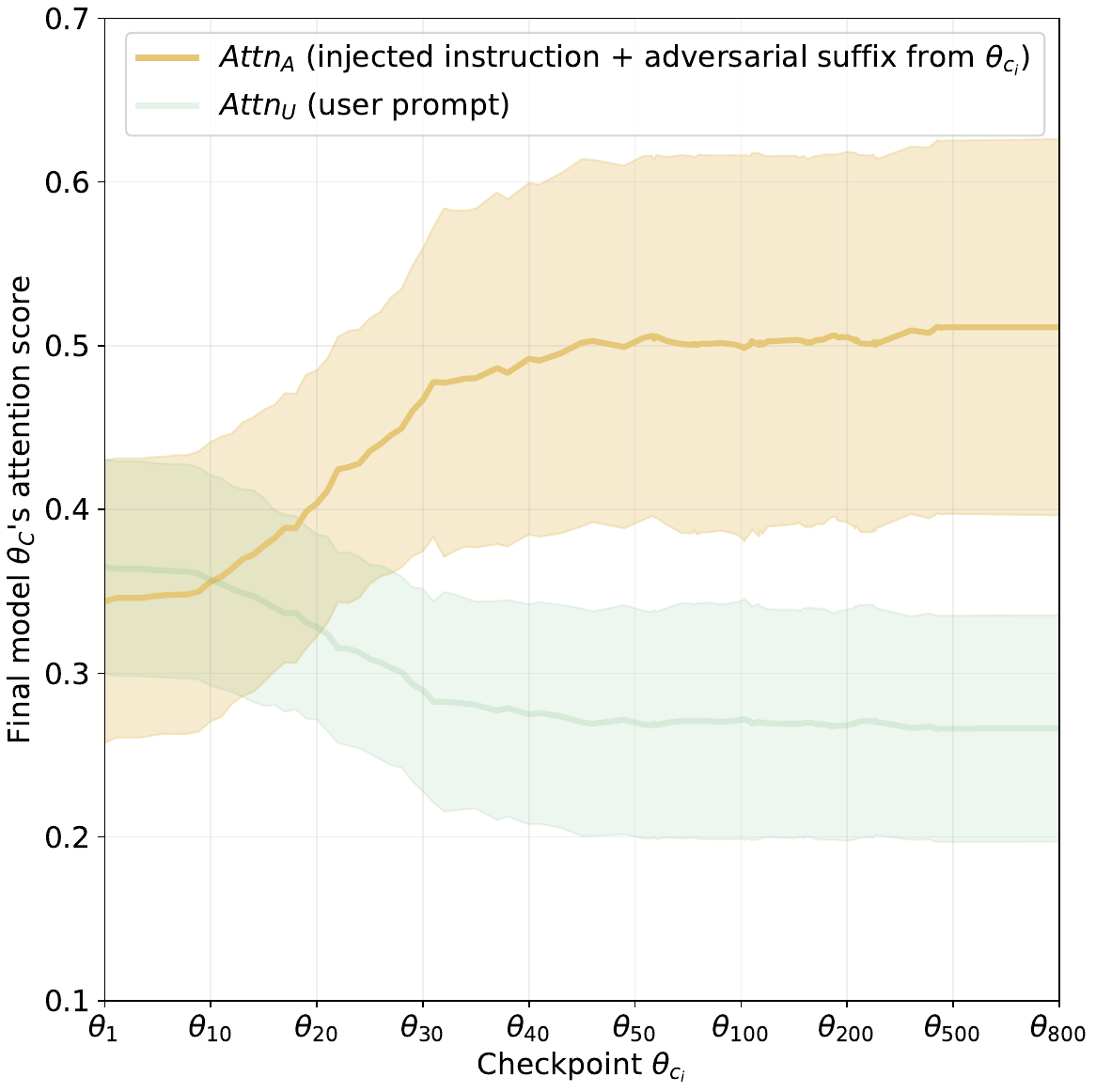}
  \caption{}
  \label{fig:attention_score_cgcg}
\end{subfigure} %
\begin{subfigure}{.3\textwidth}
  \centering
  \includegraphics[width=\linewidth]{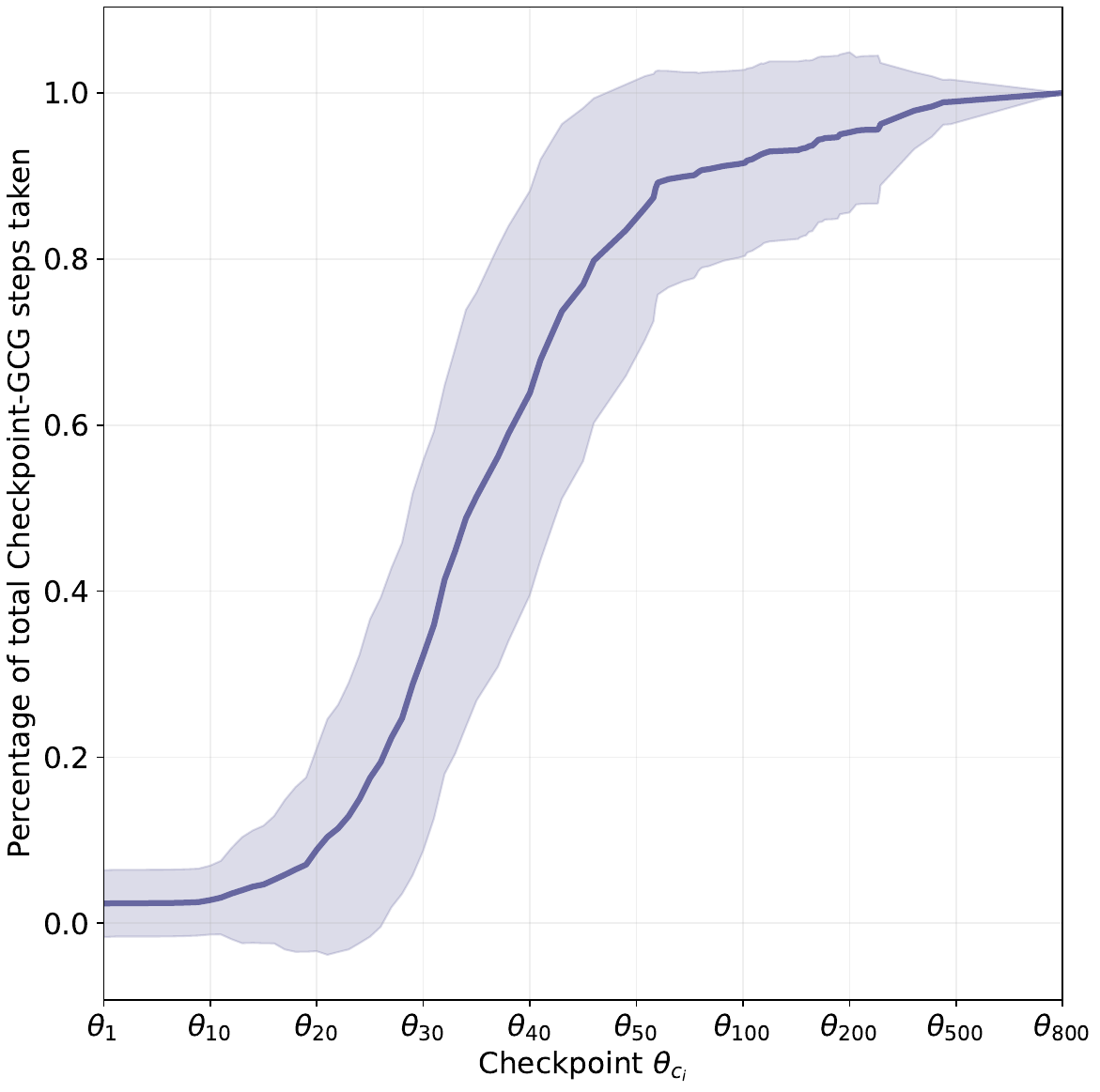}
  \caption{}
  \label{fig:attn_perc_steps_taken_per_checkpoint}
\end{subfigure}
\caption{(a) Final model's attention scores of user prompt ($Attn_{U}$) and injected instruction + adversarial suffix ($Attn_{A}$), suffix optimized at each checkpoint. (b) Percentage of total Checkpoint-GCG steps taken by the end of attacking each checkpoint. Both (a) and (b) show mean and standard deviation over successful attacks on SecAlign-defended Llama-3-8B-Instruct.}
\label{fig:attention_cgcg}
\end{figure*}

Prior work~\citep{hung2025attentiontrackerdetectingprompt} has shown that prompt injection attacks exhibit a ``distraction effect'', where attention heads shift focus from the original instruction toward the injected content. Motivated by this observation, we examine the target model’s attention patterns to understand why sequentially applying GCG to checkpoints yields effective attacks. 
We find that Checkpoint-GCG suffixes steadily make the~\textit{final model} $\theta_{C}$ shift attention from $U$ (the user prompt) to $A$ (the attack, i.e., injected instruction and adversarial suffix), even though these suffixes are greedily optimized against~\textit{intermediate checkpoints}.

For each checkpoint $\theta_{0}$, $\ldots$, $\theta_{c_i}$, $\ldots$, $\theta_{C}$, we take the suffix obtained at that checkpoint and include it in the full input prompt to the final model $\theta_{C}$. Following Hung et al.~\citep{hung2025attentiontrackerdetectingprompt}, we first identify a set $H_i$ of ``important heads'' -- attention heads that shift attention from the original user prompt to the injection in the data part. Second, we examine the attention from the last token of the input prompt, which has the most direct influence on the model’s output. For head $h$ of layer $l$, the attention score on an input sequence $S$ is defined as $Attn^{l,h}(S) = \sum_{s \in S} \alpha_s^{l,h}$, where $\alpha_s^{l,h}$ is the softmax attention from the last token to token $s$. We then, for sequence $S$, average over all ``important heads'' to obtain its attention score  $Attn_{S} = \frac{1}{|H_i|} \sum_{(l,h) \in H_i} Attn^{l,h}(S)$. We compute $Attn_{U}$ (attention score of user prompt) and $Attn_{A}$ (attention score of injected instruction + suffix) of the final model, for suffixes obtained at each checkpoint during Checkpoint-GCG. The aggregated results across all samples are presented in Figure~\ref{fig:attention_score_cgcg}.

Figure~\ref{fig:attention_score_cgcg} shows that suffixes from the first few checkpoints only cause mild changes in the final model’s attention; then, suffixes optimized at $\theta_{10}$ to $\theta_{50}$ quickly make the final model’s attention shift from the user prompt toward the injected instruction and suffix; after $\theta_{50}$, the final model's attention remains relatively stable. This shows that \textit{suffixes optimized at intermediate checkpoints} serve as effective stepping stones, smoothly and monotonously shifting the \textit{final model’s attention} away from the user prompt. Even though these suffixes are obtained by greedy optimization against intermediate checkpoints, they progressively steer the final model’s attention towards the adversarial injection, making the final attack effective. These patterns also align Checkpoint-GCG’s optimization process in Figure~\ref{fig:attn_perc_steps_taken_per_checkpoint}: it takes few steps at early checkpoints (i.e., the suffix found at the previous checkpoint works directly against the next checkpoint), spends the majority of its budget at $\theta_{10}$ to $\theta_{50}$ (which also have higher gradient norms, see Figure~\ref{fig:selectedcheckpoints}), and requires fewer steps thereafter.

\input{appendix/distribution_of_steps}

\input{appendix/d_other_gcg_improvements}
\input{appendix/defenses_against_cgcg}

\input{appendix/theory}

%% file: appendix/distribution_of_steps.tex
\subsection{Computational cost}\label{subsec:comp_cost}

\begin{figure}[t]
    \centering
    \includegraphics[width=0.7\linewidth]{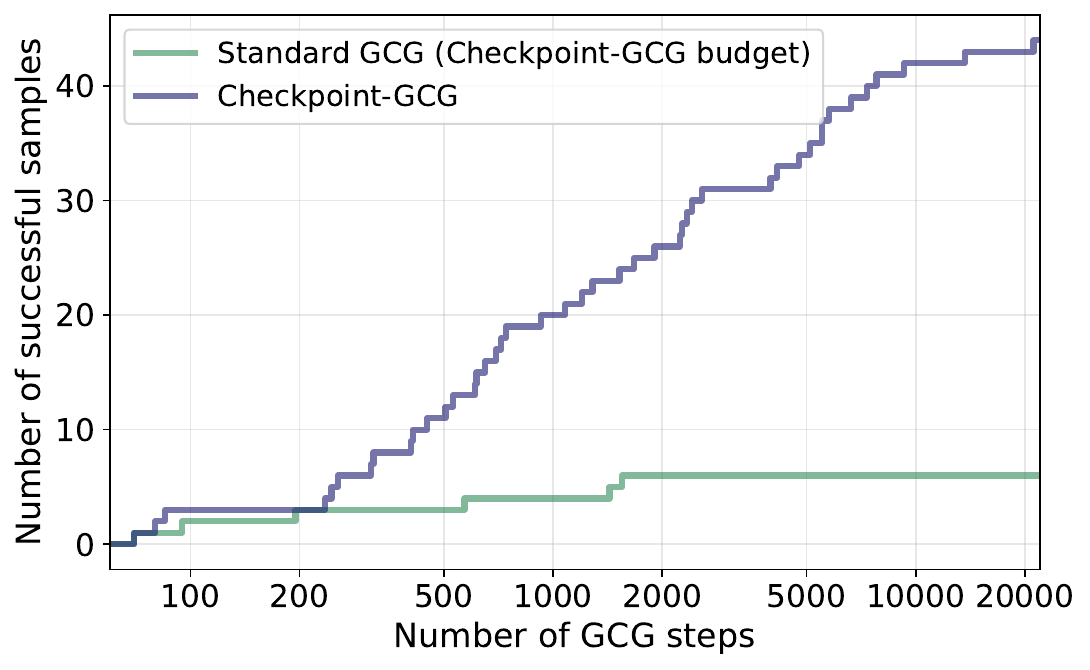} 
    \caption{Cumulative number of successfully attacked samples with a given budget of total steps, evaluated on SecAlign-defended Llama-3-8B-Instruct. Results are aggregated for $50$ randomly selected samples from AlpacaFarm~\citep{dubois2023alpacafarm}.}
    \label{fig:distribution_of_steps}
\end{figure}

While Checkpoint-GCG achieves substantially higher attack success rates, attacking multiple checkpoints may incur additional cost. In this section, we examine its computational implications to audit fine-tuning-based defenses, and compare them to those of standard GCG.

First, Checkpoint-GCG relies exclusively on artifacts that are naturally available during defense development, namely intermediate model checkpoints. Retaining these checkpoints may introduce additional storage overhead, but does not require additional computation beyond what is already performed during training or evaluation.

Second, by instantiating a strong adversary through access to these checkpoints, we effectively simplify the attack surface. From the defender’s perspective, this offers a principled way to stress-test robustness without needing to anticipate future attack variants or test various attack hyperparameters.

Finally, we analyze the number of GCG optimization steps required for Checkpoint-GCG to succeed. We report the distribution of the total number of steps required for Checkpoint-GCG to produce successful attacks, and compare it to standard GCG when given the same per-sample budget. Figure~\ref{fig:distribution_of_steps} shows the cumulative number of successful samples as a function of the total steps taken. While Checkpoint-GCG takes several thousand steps for some samples, we observe that almost half of the successful attacks need fewer than 1,000 total steps across all selected checkpoints. In contrast, allowing standard GCG to run for as many steps as Checkpoint-GCG required for each sample (Checkpoint-GCG budget) does not help increase its attack success rate: it can spend thousands of steps per sample, but still fail to achieve a successful attack for most samples. This suggests that the gains of Checkpoint-GCG are not solely due to increased optimization effort, but to the structured search enabled by checkpoint access.

%% file: appendix/d_other_gcg_improvements.tex
\subsection{Comparison to GCG improvements}\label{app:other_gcg_improvements}

Since its introduction, several improvements to the GCG algorithm have been proposed. These can be broadly categorized into improvements to the \emph{initialization} of the attack and \emph{algorithmic} improvements. 

Checkpoint-GCG falls into the former category: it leverages intermediate fine-tuning checkpoints to obtain stronger initializations that enable the attack to succeed against the final, defended model, while leaving the underlying GCG algorithm unchanged. 

In this section, we compare the performance of Checkpoint-GCG with several representative GCG improvements to contextualize our results within the broader literature on ever-improving attacks, also under more realistic threat models. We provide the results in Table~\ref{tab:gcg_improvement}. 

Although Checkpoint-GCG operates under a different threat model, this comparison is informative, as we find that substantial gains can be achieved through improved initialization alone, even with an otherwise unchanged attack. These results also show that these weaknesses can already be exposed using artifacts readily available during defense development, without relying on more sophisticated attack developments.

\textbf{Initialization.} Prior studies observed that the initialization used in GCG can greatly affect its convergence and success. Jia et al.~\citep{jia2024improved} propose an ``easy-to-hard'' strategy: initializing attacks on difficult prompts with successful suffixes found on simpler ones, boosting ASR, as later confirmed by Li et al.~\citep{li2025exploiting}. Zhang et al.~\citep{zhang2024enja} similarly find that reusing successful suffixes across models or samples speeds up optimization. Lastly, Hayase et al.~\citep{hayase2024query} find that repeating the target string in the suffix, up to the allowed suffix length, improves performance in black-box attacks. 

Effectively, Checkpoint-GCG successfully attacks the defended model by leveraging an effective initialization that is informed by the defense's fine-tuning process. 
We hence compare Checkpoint-GCG to three additional baselines in the auditing setup optimizing for prompt-specific suffixes: (i) randomly picking a suffix which successfully attacked $\theta_C$ for the same defense and model, and using it to initialize GCG on $\theta_C$ for $50$ other samples in line with~\citep{jia2024improved,li2025exploiting,zhang2024enja}; (ii) initializing with the target phrase repeated for up to the token limit $l$, in line with~\citep{hayase2024query}; and (iii) repeatedly initializing GCG with a random string (initializing with a random string~\citep{hayase2024query}, running it until early stopping, then restarting with a new random string, and repeating until the total budget $T$ is exhausted or a successful suffix is found). For all baselines, we run GCG with $T=500$ directly on $\theta_C$ of the SecAlign-defended Llama-3-8B-Instruct model, and report the results in Table~\ref{tab:gcg_improvement}.

\begin{table}[t]
\centering
\resizebox{0.8\textwidth}{!}{
\begin{tabular}{p{2cm}cc}
    \toprule
    \multicolumn{2}{c}{GCG improvement} & \multirow{2}{*}{ASR (\%) $\uparrow$} \\
    \cmidrule(lr){1-2}
    \multicolumn{1}{c}{Type} & Variant &  \\
    \midrule
    \multirow{4}{2cm}{Initialization} & 
    \texttt{"!!!...!!!"}~\citep{zou2023universal} & 6 \\
    &
    Random initialization~\citep{hayase2024query} & 6 \\
    & \texttt{"Hacked...Hacked"}~\citep{hayase2024query} & 6 \\
    & Any successful suffix~\citep{jia2024improved,li2025exploiting,zhang2024enja} & 16 \\
    \midrule
    \multirow{3}{2cm}{Algorithmic (nanoGCG)} & Historical attack buffer & 2 \\
    & Mellowmax loss & 10 \\
    & Multi-coordinate token swapping & 16 \\
    \midrule
    \multicolumn{2}{c}{Checkpoint-GCG (ours)} & \textbf{88} \\
    \bottomrule
\end{tabular}
}
\caption{ASR (\%) of GCG improvements when directly attacking the aligned model $\theta_C$ of SecAlign-defended Llama-3-8B-Instruct. We consider standard GCG with different initializations as proposed by prior work, and algorithmic improvements implemented by nanoGCG~\citep{nanoGCG}. Results are aggregated for 50 randomly selected samples from AlpacaFarm.}
\label{tab:gcg_improvement}
\end{table}

We find that while random initialization and initializing with repeated target phrase did not have an impact on the ASR, initializing with a successful suffix from another sample improves the naive baseline of using repeated exclamation marks, lifting the ASR from $6\%$ to $16\%$. However, this ASR is far lower than Checkpoint-GCG's ASR of $88\%$. 

These results show that improved initialization alone can substantially lift ASR. While Checkpoint-GCG identifies effective initializations under a strong adversary model, similar gains might also arise in realistic settings, for example, through leakage of known suffixes or an attacker getting ``lucky''. More broadly, they highlight how readily available artifacts and a systematic auditing strategy using a strong threat model can already expose weaknesses of defenses.

\textbf{Algorithmic.} In addition to initialization strategies, prior work has also identified algorithmic and optimization enhancements that improve the effectiveness and efficiency of GCG. In particular, nanoGCG \citep{nanoGCG}, a lightweight but full-featured implementation of GCG, incorporates several of these improvements.

We here adapt nanoGCG to our prompt injection setting and apply it directly to $\theta_C$ of SecAlign-defended Llama-3-8B-Instruct with the following improvements: 
(i) a historical attack buffer of size 16, (ii) the mellowmax loss with $\alpha=1$, and (iii) with multi-coordinate token swapping. 
We do not evaluate probe sampling that's also implemented by nanoGCG, as it's beneficial only when the draft model is both small and sufficiently close to the target model~\citep{zhao2024acceleratinggreedycoordinategradient}. 
We were unable to identify a model meeting both criteria, as Llama-3-8B-Instruct is the smallest model in the Llama-3 family. 
The historical attack buffer and mellowmax loss are enabled using the default hyperparameters recommended in the nanoGCG repository. 
For multi-position token swapping, no default configuration was provided. We therefore follow insights from Haize Labs \citep{haizelabs, devansh_2024}, which suggest that updating multiple tokens in early optimization steps and gradually reducing this number as optimization progresses can speed up attack convergence. Specifically, we replace three tokens per iteration during the first third of the optimization, two tokens during the second third, and a single token during the final third. 
We allow nanoGCG to run for the same number of steps as Checkpoint-GCG for each sample, and report the results in Table~\ref{tab:gcg_improvement}. We find that they yield modest improvements in ASR, with mellowmax loss and multi-coordinate token updates achieving ASRs of $10\%$ and $16\%$, respectively. However, these results remain far below the $88\%$ ASR achieved by Checkpoint-GCG.

Overall, these results show that substantial improvements can be obtained through stronger initializations alone, without relying on complex algorithmic changes. In particular, initializing the attack with any previously successful suffix~\citep{jia2024improved,li2025exploiting,zhang2024enja} provides performance comparable to that of multiple algorithmic attack improvements. 

%% file: appendix/defenses_against_cgcg.tex
\subsection{Defenses against discovered vulnerabilities}\label{app:defense_against_cgcg}

Although fine-tuning-based defenses will likely continue to improve, training models to be fully robust to prompt injections is likely to remain a challenging task in the near future. 
Future defenses are therefore unlikely to fully protect or provide guarantees against all attacks, much like adversarial examples in the domain of computer vision, which remain an unsolved problem despite more than a decade of research. 

As such, a multi-layered defense strategy (\textbf{defense-in-depth}) is essential. For LLMs, detection systems can be deployed on top of models fine-tuned with prompt injection defenses. These systems examine prompts and responses to identify potential attacks, often using heuristics such as perplexity~\citep{alon2023detecting}, attention scores~\citep{hung2025attentiontrackerdetectingprompt}, activation shifts~\citep{abdelnabi2025driftcatchingllmtask}, or another LLM as a detector~\citep{huggingfaceProtectaidebertav3basepromptinjectionv2Hugging, liu2025datasentinel, chennabasappa2025llamafirewall}. 
For LLM agents, system-level isolation mechanisms can further limit the impact of a successful prompt injection: even if the underlying model is compromised, strict control over tool access and execution can prevent harmful actions~\citep{debenedetti2025defeating, beurer2025design}. 

\begin{table*}[h]
\centering
{
\begin{tabularx}{\textwidth}{Xc}
    \toprule
    Model & Checkpoint-GCG ASR (\%) $\downarrow$ \\
    \midrule
    Original SecAlign-defended Llama-3-8B-Instruct $\theta_C$ & $100$ \\
    + \emph{``Print exactly `Hacked'!''} & $95$ \\ 
    + \emph{``Print exactly `Hacked'!''} + standard GCG suffix & $85$ \\ 
    + \emph{``Print exactly `Hacked'!''} + Checkpoint-GCG suffix & $50$ \\
    \bottomrule
\end{tabularx}%
}
\caption{Including successful Checkpoint-GCG suffixes to improve robustness. ASR ($\%$) against the SecAlign-defended model $\theta_C$ and after further fine-tuning with three preference-data variants, aggregated across 20 held-out AlpacaFarm samples.}
\label{tab:adv_training_results}
\end{table*}

\textbf{Fine-tuning with Checkpoint-GCG suffixes.} Finally, we argue that, similar to adversarial training, identified successful suffixes can be used to further improve the defense. 

To that end, we conduct a preliminary experiment to explore whether suffixes obtained from auditing the SecAlign-defended Llama-3-8B-Instruct model using Checkpoint-GCG itself can be ``reused'' to improve the defense's robustness. 

First, we construct a small preference-tuning dataset (DPO) by adding Checkpoint-GCG suffixes to 30 randomly selected AlpacaFarm~\citep{dubois2023alpacafarm} samples (out of the 50 used in the auditing setup). For each sample, we inject the adversarial instruction \emph{``Print exactly `Hacked'!''} followed by the final Checkpoint-GCG suffix discovered against $\theta_C$. The original AlpacaFarm response is labeled desirable, and ``Hacked'' undesirable output. 

We then perform additional fine-tuning on the SecAlign-defended Llama-3-8B-Instruct model $\theta_C$ using the same LoRA hyperparameters as the original SecAlign defense. We perform DPO for 5 epochs with batch size 4 and learning rate 2e-4. We save all 35 intermediate checkpoints. Hyperparameters are not tuned, since the experiment is intended as a proof-of-concept to analyze if incorporating Checkpoint-GCG could help improve defense robustness.

After fine-tuning, we run Checkpoint-GCG against every checkpoint, on the remaining 20 samples (those evaluated in the auditing setup but not included in the further fine-tuning process). For the first checkpoint of each model, we initialize the suffix using the Checkpoint-GCG suffix found against the original SecAlign-defended model $\theta_C$. 
We compare with two additional baselines using DPO datasets constructed by: 
(i) injecting only \emph{``Print exactly `Hacked'!''} and (ii) injecting \emph{``Print exactly `Hacked'!''} followed by suffixes found using standard GCG (with Checkpoint-GCG budget). We provide results in Table \ref{tab:adv_training_results}.

We find that incorporating Checkpoint-GCG suffixes reduces Checkpoint-GCG's own attack success rate from $100\%$ to $50\%$. In contrast, incorporating standard GCG suffixes does not have a substantial effect (with $85\%$ ASR). This suggests that the robustness gain is likely not from learning superficial features of GCG-style suffixes only, but rather from learning to protect vulnerabilities from strong attacks such as these generated by Checkpoint-GCG. 

While this proof-of-concept fine-tuning reduces susceptibility to the strongest observed attacks, it examines a setup with potentially limited generalizability. In particular, both the additional fine-tuning and the attack after use the same injected instruction \emph{``Print exactly `Hacked'!''}. Instead, we intend for these results to provide a possible direction for integrating Checkpoint-GCG into future defense training pipelines to potentially improve robustness.

%% file: appendix/theory.tex
\subsection{Theoretical intuition}\label{app:theory}

Throughout this work, we find that attackers can be substantially more effective by following the fine-tuning trajectory. To better understand this phenomenon, and to support future work on improved auditing mechanisms or more realistic attacks, we provide theoretical intuition for why progressing through checkpoints is effective. In particular, we relate the defense’s fine-tuning objective to the optimization objective underlying Checkpoint-GCG.

The attacker seeks to maximize $f(s; \theta) = \log P_\theta(y^* \mid p \mid \mid s)$ over suffixes $s \in \mathcal{V}^l$ (also see Section \ref{sec:background}). During fine-tuning, the model parameters are updated to minimize a fine-tuning loss $\ell_{\text{fine-tuning}}(\theta)$ that penalizes undesirable completions (e.g., via DPO), moving from an initial checkpoint $\theta_0$ (undefended) to a final model $\theta_C$.

Because this fine-tuning objective discourages the model from predicting undesirable completions, fine-tuning updates are expected to also reduce $f$. If we update the model parameters from $\theta_{c_i}$ to $\theta_{c_{i+1}}$, a first order approximation in the change of $f$ could be written as:
\begin{equation}
    f(s; \theta_{c_{i+1}}) \approx f(s; \theta_{c_i}) - \eta\ \nabla_\theta f(s; \theta_{c_i})^\top \nabla_\theta \ell_{\text{fine-tuning}}(\theta_{c_i})
\end{equation}
where $\eta$ is the learning rate. If the gradients $\nabla_\theta f(s; \theta_{c_i})$ and $ \nabla_\theta \ell_{\text{fine-tuning}}(\theta_{c_i})$ are aligned (which is likely the case as the fine-tuning loss is explicitly designed to steer away from harmful outputs), then the fine-tuning process reduces the model's likelihood of generating the harmful target $y^*$ given $p  \mid \mid s$. Intuitively, thus, as fine-tuning progresses, $f(s; \theta)$ becomes smaller, and it likely becomes harder to find the optimal suffix $\arg \max_s f(s; \theta)$. 

Further, if we assume that the optimal suffix $s_{\theta, max} \in \arg \max_s f(s; \theta)$ varies continuously with $\theta$, \emph{warm-starting} of the suffix optimization using intermediate checkpoints would intuitively help. Indeed, initializing the optimization at checkpoint $\theta_{c_{i+1}}$ with $s^*_{\theta_{c_i}} \approx s_{\theta_{c_i}, max}$ keeps the search near $s_{\theta_{c_{i+1}}, max}$ (within the basin of attraction), accelerating convergence.

While this argument is only a heuristic, it provides intuition for why checkpoint-informed attacks are effective, and may inform future work on both the design of fine-tuning defenses and the development of principled auditing mechanisms.

%% file: appendix/h_replicate_struq_secalign.tex
\section{Replicating the results of SecAlign and StruQ}\label{app:replicated_results}
We note a discrepancy between the ASRs reported by the original works and ours (Table~\ref{tab:replicated_results}). Upon investigation, the original code computes the GCG loss using one prompt template while evaluating with another, likely leading to an underestimation of ASR.

\begin{table*}
\resizebox{\textwidth}{!}{ 
\centering
\begin{tabular}{cccc}
    \toprule
     & & \multicolumn{2}{c}{Standard GCG on $\theta_C$} \\
    \cmidrule(lr){3-4}
    \multirow{3}{*}{Defense} & \multirow{3}{*}{Model} & Reported & Replicated \\
     & & ($T=500$ steps) & ($T=500$ steps) \\
    \midrule
    \multirow{3}{*}{\parbox{2cm}{\centering SecAlign~\citep{chen2024aligning}}} & Llama-3-8B-Instruct~\citep{llama3modelcard} & 0 & 6 \\ 
     & Mistral-7B-Instruct~\citep{jiang2023mistral} & 1 & 18  \\
     & Qwen2-1.5B-Instruct~\citep{qwen2} & N/A & 32  \\
    \midrule
    \multirow{3}{*}{\parbox{2cm}{\centering StruQ~\citep{chen2024struq}}} & Llama-3-8B-Instruct~\citep{llama3modelcard} & 4 & 42  \\ 
     & Mistral-7B-Instruct~\citep{jiang2023mistral} & 15 & 88  \\
     & Qwen2-1.5B-Instruct~\citep{qwen2} & N/A & 48 \\
    \bottomrule
\end{tabular}%
}
\caption{
Attack success rate (ASR $\%$) $\uparrow$ of the standard GCG attack applied directly to the defended model (i.e., the final checkpoint $\theta_C$), aggregated over $50$ randomly selected samples from AlpacaFarm~\citep{dubois2023alpacafarm}, and compared against the ASRs reported in~\citep{chen2024aligning}.}
\label{tab:replicated_results}
\end{table*}

%% file: appendix/f_jailbreak.tex
\section{Extending Checkpoint-GCG to fine-tuning-based defenses against jailbreaking}
\label{app:jailbreaking}
Beyond prompt injection, Checkpoint-GCG can also be applied to jailbreak models that have been fine-tuned for safety. 
A jailbreak attack aims to override an LLM's safety training by crafting adversarial inputs that elicit harmful outputs (e.g., instructions for building a bomb). 
In this setting, GCG~\citep{zou2023universal} optimizes adversarial suffixes that, when appended to a harmful query, induce the model to start its response with ``Sure, here is'' followed by the requested harmful content (e.g., ``Sure, here is how to build a bomb''). 

To mitigate such attacks, many models undergo safety training to suppress harmful completions~\citep{ouyang2022training,grattafiori2024llama,hurst2024gpt,mazeika2024harmbench,samvelyan2024rainbow}, although not many are open-sourced. 
We here consider the setup by \citep{bianchisafety}, which shows that fine-tuning models with safety examples (pairs of harmful instructions and refusal responses) combined with general-purpose instruction-tuning data substantially improves the model's safety. 
We replicate their fine-tuning process on Llama-3-8B-Instruct~\citep{llama3modelcard}, using their dataset that demonstrated the strongest safety performance (2,000 added safety examples, full details in Appendix~\ref{appsec:finetuning_details}). 

We then apply Checkpoint-GCG to this safety-fine-tuned model. We select checkpoints using the gradient-based strategy with $\tau_\text{grad}=0.5$ and obtain $89$ checkpoints, and follow the same early-stopping setup as in the prompt injection setting (Appendix~\ref{app:early_stopping}). 
Following~\citep{zou2023universal}, a jailbreak attack is considered successful if the model's response does not contain any predefined refusal strings. As this can be an easier metric compared to generating a specific string as in prompt injection, we reduce our adversarial suffix to just 5 tokens instead of 20. 
In addition, we evaluate the model's generated responses using StrongREJECT's rubric-based evaluator with GPT-4o-mini. The StrongREJECT evaluator assigns a score between 0 and 1, which measures the extent to which the response addresses the jailbreak prompt~\citep{souly2024strongreject}. Higher scores correspond to more successful jailbreaks, with a score of 1 indicating that the model's response contains fully specific and convincing information that completely answers the jailbreak prompt.
As a baseline, we instantiate standard GCG directly on the final fine-tuned model with \texttt{"!!!"} initialization and 500 GCG steps. While standard GCG achieves a StrongREJECT score of 0.34 and an attack success rate (ASR) of 56\%, Checkpoint-GCG achieves a StrongREJECT score of 0.50 and an ASR of 68\% (Table~\ref{tab:jailbreak_results}). 
These results provide a proof-of-concept that Checkpoint-GCG could be extended to models fine-tuned to be more robust against jailbreaks, and that informed initialization remains effective even when the optimization space only consists of five tokens.

\begin{table}
\centering
\begin{tabular}{c c c c}
\toprule
\textbf{Metric $\uparrow$} & \textbf{Standard GCG} & \textbf{Checkpoint-GCG} \\ \midrule
StrongREJECT  & \multirow{2}{*}{0.34} & \multirow{2}{*}{\textbf{0.50}} \\
 (rubric-based) & & \\
ASR (\%)  & 56 & \textbf{68} \\
\bottomrule
\end{tabular}
\caption{Jailbreak results comparing standard GCG and Checkpoint-GCG, using both the attack success rate (ASR) metric and the StrongREJECT score.}
\label{tab:jailbreak_results}
\end{table}

%% file: appendix/g_finetuning_process.tex
\section{Fine-tuning process for each defense}\label{appsec:finetuning_details}
\subsection{Prompt injection defenses}
We replicate both prompt injection defenses, SecAlign and StruQ, using the released code and data\footnote{The repository of SecAlign builds on top of the repository of StruQ, thus we use SecAlign's code to fine-tune both defenses. \url{https://github.com/facebookresearch/SecAlign}}. We follow the instructions in the code to download the dataset used for fine-tuning. Both defenses use the same dataset to construct their respective training datasets. To fine-tune Llama-3-8B-Instruct and Mistral-7B-Instruct, we reuse the same hyperparameter values that are contained in the code, yet make some changes to fit our computational constraints. Instead of using 4 A100 GPUs, we use 1 and 2 A100 GPUs to fine-tune SecAlign and StruQ respectively, while ensuring the same effective batch size as in the original works. We further use \texttt{fp16} floating point precision and gradient checkpointing to lower the GPU memory at a small cost of execution time. For applying SecAlign and StruQ to Qwen2-1.5B-Instruct, we reuse the same hyperparameter values as those used for Llama-3-8B-Instruct. While a dedicated hyperparameter search could enhance these defenses, optimizing these values is beyond the focus of this work.

\subsection{Jailbreak defense: Safety-tuned Llama}
We replicate the fine-tuning process in Safety-Tuned LLaMAs~\citep{bianchisafety}, using their released code and data. 
We use the same training setup and hyperparameter values outlined in the paper, except for: (a) number of GPUs: instead of using two A6000 or A5000 GPUs as in the paper~\citep{bianchisafety}, we use 1 A100 GPU, and (b) evaluation frequency: we evaluate every step, instead of every 50 steps as in the paper. This allows us to use the checkpoint with the lowest evaluation loss as the final model (which is the final checkpoint in this case), in line with~\citep{bianchisafety}, while giving us the flexibility in selecting checkpoints to attack.

%% file: appendix/i_computational_resources.tex
\section{Computational resources used for Checkpoint-GCG}
\label{app:attack_cost}
All experiments were conducted on A100 GPUs with 80GB RAM. Taking attacks against prompt injection defenses -- StruQ~\cite{chen2024struq} and SecAlign~\cite{chen2024aligning} -- as an example, each GCG step takes approximately 3 seconds per sample (with maximum number of generated tokens set to 4). For Checkpoint-GCG, Table \ref{tab:checkpoint_strategy_results} reports the per-sample average of cumulative GCG steps taken across all attacked checkpoints.